\documentclass[runningheads,citeauthoryear]{llncs}

\usepackage{iftex}
\ifPDFTeX
	\usepackage[utf8]{inputenc}
	\DeclareUnicodeCharacter{0301}{\'{}} %
\fi

\newcommand{\ifAnon}[2]{#2}

\usepackage[disable]{todonotes}

\usepackage{graphicx} %
\usepackage{amsmath}  %
\usepackage{amssymb}  %
\usepackage{booktabs} %
\usepackage{subcaption} %
\usepackage{adjustbox}
\usepackage{color}
\usepackage{colortbl}
\usepackage{csquotes}
\usepackage{xfrac}
\usepackage{xspace}
\usepackage{colortbl}
\usepackage{mathtools}
\usepackage{stmaryrd}
\usepackage[
	pdfencoding=auto,
	psdextra,
	colorlinks=true,
	linkcolor=blue!50!black,
	citecolor=blue!50!black,
	urlcolor=blue!50!black,
	bookmarks=true,
	bookmarksopen=true,
	bookmarksnumbered=true,
	pdfstartview=FitH,
	bookmarksdepth=2 %
]{hyperref}
\usepackage[nameinlink]{cleveref}
\renewcommand{\cref}{\Cref}
\usepackage{thm-restate}
\usepackage{multirow}
\usepackage{braket}
\usepackage{siunitx}
\usepackage{tikz}
\usetikzlibrary{positioning,shapes.geometric,arrows,bending,decorations.pathreplacing,calc,fit}
\usepackage{pgfplots}
\pgfplotsset{compat=newest}
\usepgfplotslibrary{groupplots}
\usepackage[normalem]{ulem}
\usepackage[most]{tcolorbox}
\usepackage{nicefrac}
\usepackage{dirtytalk}
\usepackage{transparent}
\usepackage{wrapfig}
\usepackage[inline]{enumitem}
\usepackage{listings}
\usepackage{array}
\usepackage{orcidlink}
\usepackage{marvosym} %

\usepackage[
  backend=biber,
  style=lncs
]{biblatex}
\DeclareSourcemap{
  \maps[datatype=bibtex]{
    \map{
      \step[fieldsource=url, match=\regexp{\\_}, replace=_]
    }
    \map{
      \step[fieldsource=doi, match=\regexp{\\_}, replace=_]
    }
  }
}

\providecommand{\Description}[1]{}

\makeatletter
\providecommand{\leftsquigarrow}{%
  \mathrel{\mathpalette\reflect@squig\relax}%
}
\newcommand{\reflect@squig}[2]{%
  \reflectbox{$\m@th#1\rightsquigarrow$}%
}
\makeatother

\makeatletter
\let\c@lemma=\c@theorem
\let\c@definition=\c@theorem
\let\c@example=\c@theorem
\makeatother

\definecolor{hintgray}{RGB}{92,92,92}

\definecolor{col1}{RGB}{254,195,10}
\definecolor{col2}{RGB}{251,0,6}
\definecolor{col3}{RGB}{252,11,135}
\definecolor{col4}{RGB}{17,139,2}
\definecolor{col5}{RGB}{14,131,254}
\definecolor{col6}{RGB}{82,0,135}

\definecolor{brLightGreen}{HTML}{addd8e}
\definecolor{brLightGray}{HTML}{f0f0f0}

\definecolor{orange}{RGB}{230,159,0}
\definecolor{skyblue}{RGB}{86,180,233}
\definecolor{brBlue}{RGB}{0,114,178}
\definecolor{bluishgreen}{RGB}{0,158,115}
\definecolor{vermillion}{RGB}{213,94,0}
\definecolor{reddishpurple}{RGB}{204,121,167}

\colorlet{heyvlColor}{vermillion!60!black}
\definecolor{prepostColor}{RGB}{0,69,107}
\colorlet{stmtColor}{bluishgreen!50!black}

\newcommand{\toolname}[1]{\textsc{#1}}

\newcommand{\sfsymbol}[1]{\textsf{\upshape {#1}}}

\newcommand{\HeyVL}{\sfsymbol{HeyVL}\xspace}
\newcommand{\HeyLo}{\sfsymbol{HeyLo}\xspace}

\newcommand{\Caesar}{\sfsymbol{Caesar}\xspace}

\newcommand{\PosRealsInf}{\mathbb{R}_{\geq 0}^{\infty}}

\newcommand{\symStmt}[1]{\ensuremath{{\color{stmtColor}{#1}}}}

\newcommand{\symIf}{\symStmt{\texttt{if}}}

\newcommand{\symAssert}{\symStmt{\texttt{assert}}}
\newcommand{\symAssume}{\symStmt{\texttt{assume}}}

\newcommand{\symWhile}{\symStmt{\texttt{while}}}

\newcommand{\symProc}{\symStmt{\texttt{proc}}}
\newcommand{\symcoProc}{\symStmt{\texttt{coproc}}}

\newcommand{\symPre}{\symStmt{\texttt{pre}}}
\newcommand{\symPost}{\symStmt{\texttt{post}}}

\newcommand{\symAnnotate}{\symStmt{\texttt{@}}}

\newcommand{\symDomain}{\symStmt{\texttt{domain}}}

\newcommand{\stmtAnnotate}[3]{\ensuremath{\symAnnotate{}{\symStmt{\texttt{#1}}}({#2})~{#3}}}

\newcommand{\symWp}{\sfsymbol{wp}}
\newcommand{\symVc}{\sfsymbol{vp}}

\newcommand{\symErt}{\sfsymbol{ert}}

\newcommand{\approxUnder}{\textsf{Under}\xspace}
\newcommand{\approxOver}{\textsf{Over}\xspace}

\newcommand{\lowerBound}{\textsf{LB}\xspace}
\newcommand{\upperBound}{\textsf{UB}\xspace}

\definecolor{DodgerBlue3}{RGB}{24,116,205}

\lstdefinelanguage{HeyVL}{
	morekeywords={var,while,if,else,@invariant,@slice_verify,@slice_error,Int,Bool,EUReal,UReal,UInt,proc,coproc,pre,post,flip,coassume,assume,assert,coassert,\cup,cohavoc,havoc,domain,func,axiom,@wp},
	morecomment=[l]{//},
	sensitive=true
}

\definecolor{darkred}{rgb}{0.8, 0.0, 0.0}
\newcommand{\marknewtext}[1]{#1}

\begin{document}

\title{\textsc{Caesar}: A Deductive Verifier\\
	for Probabilistic Programs}
\titlerunning{\textsc{Caesar}}

\author{Philipp Schröer$^{(\text{\Letter})}$\inst{1}\orcidlink{0000-0002-4329-530X} \and
	Kevin Batz\inst{2}\orcidlink{0000-0001-8705-2564}\and
	Umut Yiğit Dural\inst{1}\orcidlink{0009-0002-8071-3600}\and
	Darion Haase\inst{1}\orcidlink{0000-0001-5664-6773} \and
	Benjamin Lucien Kaminski\inst{4,5}\orcidlink{0000-0001-5185-2324}\and
	Joost-Pieter Katoen\inst{1}\orcidlink{0000-0002-6143-1926}\and
	Christoph Matheja\inst{3,6}\orcidlink{0000-0001-9151-0441}
}%
\authorrunning{P. Schröer et al.}
\institute{%
	RWTH Aachen University, Germany \\
	\email{\{phisch, darion.haase, katoen\}@cs.rwth-aachen.de\\umut.dural@rwth-aachen.de}
	\and
	Cornell University, USA \\
	\email{k.batz@ucl.ac.uk}
	\and
	DTU Compute, Denmark \\
	\and
	Saarland University, Germany \\
	\email{kaminski@cs.uni-saarland.de}
	\and
	University College London, UK
	\and
	University of Oldenburg, Germany \\
	\email{christoph.matheja@uol.de}
}

\maketitle

\begin{abstract}
	\emph{\Caesar} is a deductive verifier for probabilistic programs.
	At its core lies \HeyVL, a quantitative \emph{intermediate verification language} based on the real-valued logic \HeyLo.
	\HeyVL allows users to express a probabilistic program, its specifications, and proof rules in a programming-language style, so that new proof rules can be easily integrated into the verifier.
	\Caesar translates \HeyVL programs into verification conditions, which are then checked using the Z3 SMT solver.
	It also includes a backend based on probabilistic model checking for a subset of \HeyVL.
	We report on the results of five years of development of \Caesar, highlighting its main features and architecture.
	In particular, we describe recent improvements such as additional proof rules, a model-checking backend, and better diagnostics.
\end{abstract}

\section{Introduction}

We present \Caesar\footnote{\url{https://www.caesarverifier.org}}, an open-source deductive verifier for probabilistic programs, written in Rust.
It is designed as verification infrastructure for a variety of properties of probabilistic programs.

\paragraph{Probabilistic programs.}
Probabilistic programs incorporate random choices whose outcomes can influence the control flow of loops and conditionals.
For example,~\Cref{fig:heyvl-brp-send-failures-upper-bound} shows the bounded retransmission protocol~\cite{DBLP:conf/tacas/DArgenioKRT97} (BRP) with a noisy transmission channel, repeatedly trying to submit a number of packets with \texttt{maxTries} attempts per packet before aborting.
The program is written in \HeyVL, a quantitative verification-aware programming language, \mbox{explained below}.

Design and verification objectives for probabilistic programs go beyond classical safety and liveness, \marknewtext{and are quantitative rather than Boolean}.
For the BRP, one may bound the failure probability, lower-bound successful transmissions, or analyze expected runtime.
\Cref{fig:heyvl-brp-send-failures-upper-bound} checks the first: it specifies that \texttt{fail == maxTries} holds on termination with probability at most $1 - (1 - 0.01^\texttt{maxTries})^\texttt{packets}$.
The \symcoProc{} keyword requests verification of an upper bound (given by the \symPre{}) on the expected value of the \symPost{}\marknewtext{expectation. Here, this \symPost{} is the condition \texttt{fail == maxTries}.
    To reason about the expected number of failures instead, one would use the \symPost{} \texttt{fail}.}
To obtain the expected value, we use the \emph{weakest pre-expectation semantics}~\cite{mciverAbstractionRefinementProof2005}, selected with the $\symStmt{\texttt{@wp}}$ annotation.
The \symWhile-loop needs a proof rule; here \emph{Park induction} (\stmtAnnotate{invariant}{\dots}{\!\!}).

\begin{figure}[t]
\begin{lstlisting}[language=HeyVL]
@wp coproc brp(packets: UInt, maxTries: UInt)
        -> (sent: UInt, fail: UInt, totalFail: UInt) 
    pre 1 - exp(1 - exp(0.01, maxTries), packets)
    post [fail == maxTries]
{
  sent = 0; fail = 0; totalFail = 0            
  @invariant(...)
  while (sent < packets && failed < maxTries) {
    var transmit: Bool = flip(0.99)
    if transmit {
      sent = sent + 1; failed = 0
    } else {
      fail = fail + 1; totalFail = totalFail + 1 } } }
\end{lstlisting}
    \Description{Excerpt of a HeyVL \symcoProc{}edure \texttt{brp} modeling the bounded retransmission protocol.
        \texttt{brp} takes \texttt{packets} and \texttt{maxTries} as inputs and outputs \texttt{sent}, \texttt{fail}, and \texttt{totalFail}.
        A \symPost{} specifies the condition \texttt{fail == maxTries} and a \symPre{} defines the upper bound on the expected value of the \symPost{} that is to be verified as $1 - (1 - 0.01^\texttt{maxTries})^\texttt{packets}$.
        The program consists of a loop that sends packets while the number of \texttt{sent} packets is less than the total \texttt{packets} to transmit and \texttt{failures} of a single packet are below \texttt{maxTries}, using a probabilistic flip (with success probability $0.99$) to decide if a packet is transmitted or counted as failed, updating counters accordingly.
        An annotation for a loop invariant is present, but the invariant itself is omitted.}
    \caption{Excerpt of \HeyVL code modeling the bounded retransmission protocol.}
    \label{fig:heyvl-brp-send-failures-upper-bound}
\end{figure}

\paragraph{The Intermediate Verification Language HeyVL.}
To tackle the diversity of verification objectives and proof rules, \Caesar uses the \emph{intermediate verification language} (IVL) \HeyVL~\cite{DBLP:journals/pacmpl/SchroerBKKM23}.
Similar to other IVLs such as \textsc{Boogie}~\cite{leino2008boogie}, \textsc{Why3}~\cite{DBLP:conf/esop/FilliatreP13}, or \textsc{Viper}~\cite{viper}, it serves as a common language to express programs, specifications, and proof rules.
\HeyVL uses the real-valued logic \HeyLo to express quantitative specifications and verification conditions.
Quantitative verification statements such as $\symAssert$ and $\symAssume$ generalize classical Hoare-style assertions and assumptions to the setting of expected values.

\begin{figure}[t]
    \centering
    \scalebox{0.88}{\begin{tikzpicture}
            \node[draw = black,thick] (program) {\begin{tabular}{c}Probabilistic\\Program\end{tabular}};
            \node[draw = black,thick, below =1em of program] (specs) {Specification};
            \node[draw = black,thick, below =1em of specs] (proofrules) {Proof Rules};

            \node[draw = vermillion,thick, right = of specs, xshift=-0.5em, yshift=-1em] (heyvl) {$\HeyVL$};
            \node[draw = prepostColor,thick, above right = of heyvl,yshift=-2em] (heylo) {$\HeyLo$};

            \node[draw = bluishgreen,thick,fill=white, right = of heylo] (smtlib) {SMT-LIB};
            \node[draw = bluishgreen,thick, right = of smtlib] (z3) {Z3};
            \node[draw = bluishgreen,thick, below = of smtlib] (jani) {JANI};
            \node[draw = bluishgreen,thick, right = of jani] (storm) {Storm};

            \node[text = green!50!black, right = of z3, xshift=1em, yshift=4em] (verified) {$\checkmark$ Verified};
            \node[text = red!50!black, below =1em of verified] (counterexample) {$\times$ Counterexample};
            \node[text = yellow!50!black, below =1em of counterexample] (unknown) {$?$ Unknown};
            \node[text = blue!50!black, below =1em of unknown] (expectedvalue) {Expected Values};

            \path (heyvl) edge [in=105,out=75,loop,->,thick,vermillion] node[above] {\footnotesize{}\textcolor{vermillion}{Slicing}} (heylo);
            \path (program) edge [->,thick,black] node {} (heyvl);
            \path (specs) edge [->,thick,black] node {} (heyvl);
            \path (proofrules) edge [->,thick,black] node {} (heyvl);
            \path (heyvl) edge [->,thick,prepostColor] node[above,sloped] {\footnotesize{}VCG} (heylo);
            \path (heylo) edge [in=105,out=75,loop,->,thick,prepostColor] node[above] (qe) {\footnotesize{}\textcolor{prepostColor}{\begin{tabular}{c} Quantifier \\ Elimination \end{tabular}}} (heylo);
            \path (heylo) edge [->,thick,bluishgreen, align=center] node[below] {\footnotesize{}VC \\ valid?} (smtlib);
            \path (smtlib) edge [->,thick,bluishgreen] node {} (z3);
            \path (heyvl) edge [->,thick,bluishgreen,bend right=10,dashed] node {} (jani);
            \path (jani) edge [->,thick,bluishgreen] node {} (storm);

            \node[draw=reddishpurple,fit=(heyvl)(storm)(z3)(qe)] (box) {};
            \node[anchor = north east, text=reddishpurple] at (box.north east) {\Caesar};

            \path (box) edge [->,thick,black,dashed] node {} (verified.west);
            \path (box) edge [->,thick,black,dashed] node {} (counterexample.west);
            \path (box) edge [->,thick,black,dashed] node {} (unknown.west);
            \path (box) edge [->,thick,black,dashed] node {} (expectedvalue.west);

        \end{tikzpicture}}

    \Description{Diagram showing the logical architecture of the \Caesar verifier.
        The inputs (Probabilistic Programs, Specifications, and Proof Rules) feed into HeyVL as a combined encoding.
        Caesar performs slicing on the HeyVL program and then transforms to HeyLo through verification-condition generation (VCG).
        Quantifier elimination is applied to the HeyLo formula before the corresponding VC validity condition is encoded in SMT-LIB and sent to Z3 for verification.
        Alternatively, a HeyVL-sublanguage can be translated to JANI and checked using Storm.
        Finally, arrows indicate outcomes of Caesar: successful verification, a counterexample, unknown verification status, or computed expected values.
    }
    \caption{Logical architecture of the \Caesar verifier.}
    \label{fig:caesar-architecture}
\end{figure}
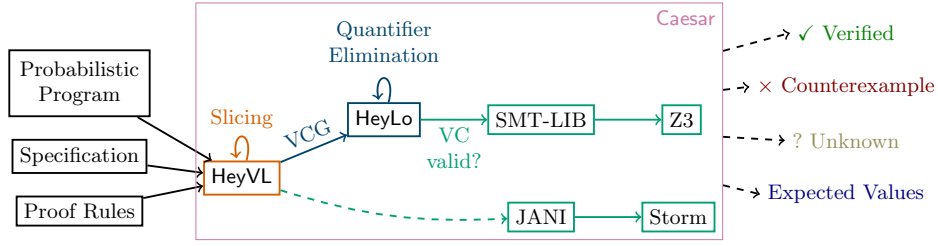

\paragraph{The Caesar Verifier.}
Main aspects of \Caesar's design are:
(1)~Probabilistic programs with $\PosRealsInf$-valued specifications, (2)~quantitative verification statements to encode a range of proof rules (e.g., inductive rules for reasoning about loops), and (3)~duality of constructs for reasoning about lower and upper bounds of expected values.
Prior work has formalised this perspective~\cite{DBLP:journals/pacmpl/SchroerBKKM23}.

\Cref{fig:caesar-architecture} shows the \emph{logical} architecture of \Caesar.
The input to \Caesar is a \HeyVL program consisting of a probabilistic program, a specification (quantitative pre- and post-conditions), and proof rules to be used during verification, e.g., loop invariants for \symWhile-loops.
Specific diagnostics about either successful or failed verification attempts are based on the \HeyVL program, and will be provided by the \emph{slicing} component (\Cref{sec:features-diagnostics}).

The \emph{verification condition generator} (VCG) generates expressions in the \HeyLo logic, whose validity implies correctness of the \HeyVL program w.r.t. its specifications.
For dispatching the validity check to a solver, an important optimization is the \emph{quantifier elimination} component which tries to remove (quantitative) quantifiers in the verification conditions.
Finally, validity is checked by translating the \HeyLo validity problem into an input to the SMT solver Z3~\cite{demouraZ3EfficientSMT2008}, following the SMT-LIB standard~\cite{BarFT-SMTLIB}.
Outputs may be either successful verification, a counterexample, or an unknown result if the solver cannot decide the validity.
We give more details on these components in \Cref{sec:backends-vcg}.

\Caesar also includes a translation from a subset of \HeyVL to the JANI modeling language~\cite{DBLP:conf/tacas/BuddeDHHJT17} for probabilistic models, which can be used to calculate expected values of probabilistic programs via the Storm model checker~\cite{DBLP:journals/sttt/HenselJKQV22}.
More details on this backend are given in \Cref{sec:backends-pmc}.

\paragraph{Contributions.}
In this tool paper, we report on the current state of \Caesar and focus on \emph{application} and \emph{implementation} aspects and features.
\marknewtext{Compared to the first publication with \Caesar~\cite{DBLP:journals/pacmpl/SchroerBKKM23},} new features presented here include additional \marknewtext{built-in} proof rules (\Cref{sec:proof-rules}), \marknewtext{support for limited functions and much improved quantifier handling (\Cref{sec:backends-vcg}),} a model checking backend via JANI/Storm (\Cref{sec:backends-pmc}), \marknewtext{a novel Visual Studio Code extension (\Cref{sec:vscode-ext}),} semantic guardrails to prevent unsound combinations of proof rules (\Cref{sec:features-guardrails}), and better diagnostics via program slicing (\Cref{sec:features-diagnostics},~\marknewtext{\cite{brutus}}).

\section{Verifying Probabilistic Programs}

\Caesar's input language \HeyVL supports different program models and verification objectives and allows specifying proof rules that should be applied to enable automated verification.

\subsection{Objectives}

\HeyVL is designed with reasoning about \emph{expected values} \marknewtext{(random variables)} of probabilistic programs as the core primitive.
Verification is structured around (co-)procedures.
Each unit of verification, either a \symProc{} or a \symcoProc{}, can be specified with quantitative analogues of pre- and post-conditions (\emph{pre} and \emph{post} for short) -- generalizing classical Hoare logic specifications.
A \symProc{} is used to verify \emph{lower bounds} (\lowerBound) on expected values and checks whether the expected value of the \symPost{} after termination is at least the \symPre's value before execution.
Dually, a \symcoProc{} verifies \emph{upper bounds} (\upperBound) on expected values, i.e., whether the expected value of the \symPost{} after termination is at most the \symPre's value before execution.

Focusing on expected values makes \HeyVL versatile, as many quantitative objectives can be reduced to expected value reasoning, including:
\begin{itemize}
    \item \emph{Probabilities} of events and \emph{cumulative distribution functions} (CDFs) can be encoded as expected values of indicator functions.
    \item \emph{Almost-sure termination} (AST) is defined as termination with probability $1$.
    \item \emph{Expected runtimes} and \emph{positive almost-sure termination} (PAST) can be handled by verifying that the expected runtime is finite.
\end{itemize}
The above includes partial- and total correctness of probabilistic programs.
\Caesar also supports reasoning about \emph{discounted} expected values, as well as quantitative reachability properties including the number of visits, time to first visit, and return times to designated locations~\cite{highlyIncremental}.
Some relational properties, like probabilistic sensitivity~\cite{DBLP:journals/pacmpl/0001BHKKM21} (the change in expected output value w.r.t.\ a change in input value), are also expressible.

\subsection{Models}

\HeyVL can express a wide range of probabilistic programs with infinite state spaces.
Program variables can be of various types, including Booleans, (un)bound\-ed integers, (extended, un)bounded reals, and lists.
User-defined types can be specified using \symDomain{} declarations with uninterpreted functions and axioms.

\HeyVL is an imperative language that supports sampling from discrete distributions, and binary (and unbounded) angelic \marknewtext{(maximizing)} and demonic \marknewtext{(minimizing)} non-determinism.
Both kinds of non-determinism can be mixed in the same program, allowing users to encode simple/two-player stochastic games~\cite{DBLP:conf/vecos/Matheja24}.
\marknewtext{Possibly unbounded loops, recursion and conditioning are not part of the core of \HeyVL, but users can analyze them via proof rules.}

\subsection{Proof Rules}\label{sec:proof-rules}

\begin{figure}[t]
    \centering
    \begin{tikzpicture}[
            font=\footnotesize,
            node distance=12mm and 16mm,
            box/.style={draw, rounded corners=2pt, align=center, inner sep=6pt, fill=blue!5},
            pill/.style={draw, rounded corners=9pt, align=center, inner sep=6pt, fill=green!10},
            bad/.style={draw, rounded corners=9pt, align=center, inner sep=6pt, fill=red!10},
            instyle/.style={box, trapezium, trapezium left angle=70, trapezium right angle=110, shape=trapezium},
            mid/.style={box}
        ]

        \node[instyle] (LB) {Lower Bound (\symProc)};
        \node[instyle, below=2em of LB] (UB) {Upper Bound (\symcoProc)};

        \node[pill, right=10em of LB] (under) {\approxUnder-approximation};
        \node[pill, right=10em of UB] (over) {\approxOver-approximation};

        \draw[->, thick, bend left=10] (LB) to node[fill=white]{Verification} (under);
        \draw[->, thick, bend left=10] (LB) to node[fill=white]{Refutation} (over);
        \draw[->, thick, bend right=10] (UB) to node[fill=white]{Verification} (over);
        \draw[->, thick, bend right=10] (UB) to node[fill=white]{Refutation} (under);
    \end{tikzpicture}

    \Description{Diagram showing which approximation of the reference semantics to use in which scenario.
        Nodes for lower and upper bounds on the left are connected by arrows to under- and over-approximation nodes on the right.
        Arrows from the lower bound point to under-approximation for verification and over-approximation for refutation.
        Arrows from the upper bound point to over-approximation for verification and under-approximation for refutation.}
    \caption{Overview of which approximation of the reference semantics to use when aiming to verify/refute a lower/upper bound.}
    \label{fig:soundness-diagram}
\end{figure}
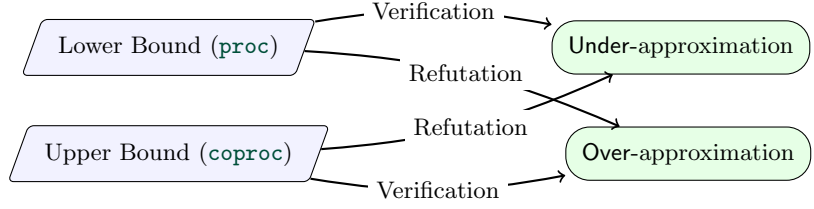

Users specify their models and objectives w.r.t.\ a reference semantics.
A reference (expectation-based) semantics for loops or recursion is defined using least or greatest fixed points (\textsf{lfp}/\textsf{gfp}) and specified using an annotation on the \symProc/\symcoProc{} (\Caesar supports \emph{weakest pre} ($\symAnnotate\symStmt{\texttt{wp}}$), \emph{weakest liberal pre} ($\symAnnotate\symStmt{\texttt{wlp}}$) (cf.~\cite{kaminskiAdvancedWeakestPrecondition2019}), \emph{expected runtimes} ($\symAnnotate\symStmt{\texttt{ert}}$)~\cite{DBLP:journals/jacm/KaminskiKMO18}, or \emph{conditional weakest pre} ($\symAnnotate\symStmt{\texttt{cwp}}$)~\cite{DBLP:journals/toplas/OlmedoGJKKM18}).

To automatically verify properties of probabilistic programs, users are required to \emph{approximate} the semantics through proof rules.
\Caesar includes a range of rules for loops that are used by adding loop annotations.
Proof rules are classified as \emph{under-approximating} (\approxUnder) or \emph{over-approximating} (\approxOver) w.r.t.\ the reference semantics.\footnote{A proof rule is \approxUnder-approximating if the expected value obtained from the proof rule is at most the expected value obtained from the reference semantics, and dually for \approxOver-approximating proof rules.}
This allows \Caesar to support both \emph{sound proofs} (verifying that a property holds) and \emph{sound refutations} (verifying that a property does not hold).
The various combinations of approximations and lower/upper bounds give rise to different soundness guarantees for proofs and refutations, as summarized in \Cref{fig:soundness-diagram}.
\marknewtext{Unsound results due to invalid combinations of proof rules are marked by \Caesar.}
\Cref{tab:proof-rules} summarizes proof rules applied with \Caesar so far.

For example, in~\Cref{fig:heyvl-brp-send-failures-upper-bound} we \approxOver-approximate the (\textsf{lfp}-based) \symWp{} semantics with the \stmtAnnotate{invariant}{\cdots}{\!\!} proof rule, i.e., the failure probability computed by \Caesar is at most the actual failure probability of the BRP.
By using the proof rule in a \symcoProc{}, the upper bound given as $\symPre$ holds for the BRP if \Caesar proves this \emph{upper bound} on the computed failure probability (\emph{sound proofs}).
Dually, if one wants to refute an upper bound, one needs to use an \approxUnder-approximating proof rule, e.g.,\ \stmtAnnotate{unroll}{\cdots}{\!\!}.
Then, a counterexample provided by \Caesar is valid for the original program too (\emph{sound refutations}).

\section{\marknewtext{Embedding Verification Techniques in \Caesar via \HeyVL}}\label{sec:ivl}
\begin{table}[t]
    \caption{Overview of proof rules in \Caesar. The syntactic sugar for built-in proof rules is given in parentheses.}
    \label{tab:proof-rules}

    \centering
    \setlength{\tabcolsep}{10pt}
    \scalebox{0.9}{
        \begin{tabular}{lll}
            \toprule
            \textbf{Semantics} & \textbf{Approx.}         & \textbf{Applicable Proof Rule}                                                                                                                                           \\
            \midrule
            lfp                & \approxOver              & Park induction~\cite{parkFixpointInductionProofs1969,kaminskiAdvancedWeakestPrecondition2019} ($\stmtAnnotate{invariant}{\dots}{\!\!}$)                                  \\
                               &                          & Latticed $k$-induction~\cite{DBLP:conf/cav/BatzCKKMS20} ($\stmtAnnotate{k\_induction}{\dots}{\!\!}$)                                                                     \\
                               &                          & Positive almost-sure termination~\cite{DBLP:conf/cav/ChakarovS13,DBLP:conf/popl/FioritiH15,kaminskiAdvancedWeakestPrecondition2019} ($\stmtAnnotate{past}{\dots}{\!\!}$) \\
                               & \approxUnder             & Finite loop unrolling~\cite{DBLP:conf/cav/BatzCKKMS20} ($\stmtAnnotate{unroll}{\dots}{\!\!}$)                                                                            \\
                               &                          & $\omega$-invariants~\cite{kaminskiAdvancedWeakestPrecondition2019} ($\stmtAnnotate{omega\_invariant}{\dots}{\!\!}$)                                                      \\
                               &                          & Optional stopping theorem~\cite{DBLP:journals/pacmpl/HarkKGK20} ($\stmtAnnotate{ost}{\dots}{\!\!}$)                                                                      \\
                               &                          & Almost-sure termination~\cite{DBLP:journals/pacmpl/McIverMKK18} ($\stmtAnnotate{ast}{\dots}{\!\!}$)                                                                      \\
                               &                          & Lower bounds for possibly divergent loops~\cite{DBLP:journals/pacmpl/FengCSKKZ23}                                                                                        \\
                               &                          & Weakly-fair AST~\cite{10.1145/3776691}                                                                                                                                   \\
            \midrule
            gfp                & \approxOver              & Finite loop unrolling~\cite{DBLP:conf/cav/BatzCKKMS20} ($\stmtAnnotate{unroll}{\dots}{\!\!}$)                                                                            \\
                               &                          & $\omega$-invariants~\cite{kaminskiAdvancedWeakestPrecondition2019} ($\stmtAnnotate{omega\_invariant}{\dots}{\!\!}$)                                                      \\
                               & \approxUnder             & Park induction~\cite{parkFixpointInductionProofs1969,kaminskiAdvancedWeakestPrecondition2019} ($\stmtAnnotate{invariant}{\dots}{\!\!}$)                                  \\
                               &                          & Latticed $k$-induction~\cite{DBLP:conf/cav/BatzCKKMS20} ($\stmtAnnotate{k\_induction}{\dots}{\!\!}$)                                                                     \\
            \midrule
            both               & \approxOver/\approxUnder & Riemann sums for continuous distributions~\cite{DBLP:journals/pacmpl/BatzKRW25}                                                                                          \\
            \bottomrule
        \end{tabular}}
\end{table}

\HeyVL{}~\cite{DBLP:journals/pacmpl/SchroerBKKM23} is a \emph{quantitative} intermediate verification language tailored to probabilistic programs.
This allows users to extend \Caesar's applicability beyond the natively supported language features by manually encoding approximations of the original semantics via verification statements.
For instance, \Caesar has been applied in this manner to programs with continuous distributions~\cite{DBLP:journals/pacmpl/BatzKRW25} and distributed randomized algorithms~\cite{10.1145/3776691}.

\subsection{HeyVL as a Quantitative Intermediate Verification Language}

\HeyVL provides \emph{verification statements} such as $\symAssert$ and $\symAssume$, which generalize classical Hoare-style assertions and assumptions to quantitative reasoning about expected values.
The core language of \HeyVL deliberately excludes loops and recursion; instead, these constructs are supported by encoding proof rules using verification statements.
This design makes the verifier extensible, enables compositional use of proof rules, and allows front-ends to benefit directly from improvements in the verification backends.

\paragraph{Expectation Transformer Semantics.}\label{sec:expectation-transformer-semantics}
\HeyVL{}'s \emph{verification pre-expectation} ($\symVc$) semantics is defined as a syntactic transformation of quantitative formulas in the real-valued logic \HeyLo.
The $\symVc$ semantics is a backwards-moving denotational semantics in the style of \emph{weakest pre-expectation} semantics~\cite{mciverAbstractionRefinementProof2005,kaminskiAdvancedWeakestPrecondition2019}.
Through quantitative assertions and assumptions \Caesar also encodes the program's specification (pre and post) in the generated formula.
The (Boolean) \emph{verification condition}, i.e., the formula representing the correctness of a \HeyVL{} program w.r.t.\ its specification, is the condition that the \HeyLo{} formula from the $\symVc$ computation is \emph{valid}.
This is the basis for \Caesar's SMT backend (\Cref{sec:backends-vcg}).

\paragraph{Operational Semantics.}
An operational semantics in terms of \emph{refereed stochastic games} has been defined for \HeyVL{}~\cite{DBLP:conf/vecos/Matheja24}.
It provides a small-step, game-based ground truth that validates the intuition behind the verification statements and proves equivalence to the weakest pre-expectation style semantics.
For a subset of \HeyVL{}, the semantics yield finite-state Markov Decision Processes (MDPs), amenable to probabilistic model checking (\Cref{sec:backends-pmc}).

\subsection{HeyLo: An Assertion Language for Probabilistic Programs}

\HeyLo{}~\cite{DBLP:journals/pacmpl/SchroerBKKM23} is a real-valued logic for the verification of probabilistic programs.
It generalizes (Boolean) first-order logic to the non-negative extended reals ($\mathbb{R}_{\geq 0}^\infty$).
We use it (1)~as a quantitative assertion language to express specifications, and (2)~to reduce many verification problems to validity checking in this logic.
For~(1), \HeyLo{} was designed as an extension of the assertion language by \citeauthor{batzRelativelyCompleteVerification2021}~\cite{batzRelativelyCompleteVerification2021}, which is relatively complete for the verification of probabilistic programs.
For~(2), \HeyLo{} is a first-order logic to simplify automation using SMT solvers, and quantitative implications and their dual coimplications allow encoding inequalities between expected values.

\section{Verification Backends}

\subsection{Verification Condition Generation}\label{sec:backends-vcg}

\Caesar's main backend is based on encoding verification conditions into \HeyLo{} and checking them using the SMT solver Z3~\cite{demouraZ3EfficientSMT2008}.
At the top level, the verification condition generator (VCG) generates a \HeyLo{} formula based on the backwards-moving $\symVc$ semantics (\Cref{sec:expectation-transformer-semantics}).

\smallskip\noindent\emph{Lazy Unfolding.}
We use a representation with explicit substitutions and shared references to sub-expressions to avoid exponential blow-up in the size of the generated formula.
\Caesar unfolds this formula lazily: it is traversed while keeping track of currently true Boolean conditions and pruning unreachable sub-expressions.\footnote{This can be seen as a form of (forwards-moving) \emph{symbolic execution}, as in KeY~\cite{DBLP:series/lncs/10001}, cf.~\cite[Section 6.3.4]{DBLP:books/mc/22/Hahnle22}. In particular, we enjoy (Boolean) forwards reasoning without, e.g., a probabilistic strongest post-expectation semantics, which has been shown to be impossible to define for probabilistic programs~\cite[p.\ 135]{DBLP:phd/ethos/Jones90}.}
The required satisfiability checks are discharged by a number of lightweight SMT queries.
The result is an equivalent \HeyLo{} formula that is often much smaller than the naively generated one.
This is an important optimization, since the generated formulas can be very large due to (non-)deterministic, and probabilistic branching in the program.

\smallskip\noindent\emph{Optimizations.}
Afterwards, \Caesar applies \emph{quantifier elimination} (QE) on \HeyLo{} formulas, where \HeyLo{} quantifiers correspond to infima and suprema over non-negative extended reals.
QE shifts quantifiers outwards as much as possible, and then eliminates top-level quantifiers while preserving validity.
Removing a quantifier introduces new fresh variables.
This procedure is not complete and may fail to eliminate some quantifiers, but is crucial for practical performance.
Fragments of \HeyLo{} and corresponding \HeyVL{} programs have been identified where QE is complete~\cite[Section 4.2]{Schroer:998370}.
Various other optimizations are implemented, such as replacing quantitative logical operators with Boolean ones where possible.

\smallskip\noindent\emph{SMT Encoding.}
The \HeyLo{} formula is translated into an input formula for Z3, making heavy use of Z3's support for real arithmetic.
To encode an expression of type $\PosRealsInf$, we represent it as a pair $(r, \textit{isInfty})$, where $r$ is a real number and $\textit{isInfty}$ is a Boolean flag that is true if and only if the represented number is equal to $\infty$.
We add constraints $r \geq 0$ to ensure that $r$ is non-negative.
All operations on $\PosRealsInf$ are then defined over such pairs.

\Caesar supports a number of different encodings for recursive uninterpreted functions declared in \symDomain{}s, which can be selected based on the verification task at hand~\cite{BeothyElo:1016969}.
We support the naive \emph{axiomatic} encoding, where axioms specify the function's behavior, as well as SMT-LIB's \texttt{define-fun-rec} definitions.
The \emph{decreasing} encoding uses \emph{patterns} on quantifiers to guide Z3's e-matching algorithm for quantifier reasoning.
Built on this, we have different variants of \emph{fuel}\marknewtext{ed function} encodings, which limit the depth of recursion during SMT solving \marknewtext{by tracking a \emph{fuel} value that bounds the allowed recursive unfoldings of the function}.
These different implementations allow users to trade off soundness of proofs and refutations, as well as completeness of verification.

\smallskip\noindent\emph{Slice Solvers.}
Diagnostics in \Caesar (\Cref{sec:features-diagnostics}) are based on computing \emph{slices} of a \HeyVL{} program~\cite{brutus}.
To this end, \Caesar implements the first semantic slicing engine for probabilistic programs through dedicated \emph{slice solvers} that operate directly on the generated \HeyLo{} verification condition.
These solvers repeatedly invoke the SMT solver to search for slices that either still verify or admit a counterexample.
\Caesar additionally supports minimization of slices.
We implement a variant of the algorithm by \citeauthor{DBLP:conf/cpaior/LiffitonM13}~\cite{DBLP:conf/cpaior/LiffitonM13} for computing \emph{minimum unsatisfiable subsets} (MUSes), sped up by exploiting polarity-like information such as whether one slices assertions (positive), assumptions (negative), or assignments (mixed).

\smallskip\noindent\emph{Completeness.}
No verifier can be complete for all probabilistic programs.
For instance, deciding almost-sure termination is known to be \emph{even more undecidable} than the Halting Problem for classical programs~\cite{DBLP:journals/acta/KaminskiKM19}.
However, we obtain specific completeness results.
For so-called \emph{linear programs} and \emph{linear expectations}, with the \stmtAnnotate{unroll}{\cdots}{\!\!}, \stmtAnnotate{invariant}{\cdots}{\!\!} and \stmtAnnotate{k\_induction}{\cdots}{\!\!} proof rules, verification is decidable for similar reasons as in~\cite[Section 7]{DBLP:conf/cav/BatzCKKMS20}.

\subsection{Probabilistic Model Checking}\label{sec:backends-pmc}

\Caesar includes a probabilistic model checking backend that compiles a subset of \HeyVL{} with loops, called \emph{executable \HeyVL}, to Markov Decision Processes (MDPs).
In contrast to the VCG backend, this translation ignores proof rule annotations and instead encodes the program's operational semantics directly as an MDP with rewards.
The resulting model is expressed in the \emph{JANI} format~\cite{DBLP:conf/tacas/BuddeDHHJT17} and expected values can be calculated with probabilistic model checkers such as \emph{Storm}~\cite{DBLP:journals/sttt/HenselJKQV22}.
Since Storm operates on finite-state MDPs, infinite-state \HeyVL{} programs require techniques like \emph{partial exploration}~\cite{DBLP:journals/theoretics/BrazdilCCFKKM0U25} to obtain lower or upper bounds on expected values.

\section{UI/UX: Design Principles and Tool Support}

Quantitative deductive verification poses distinct usability challenges: verification pre-expectations are non-Boolean, failures may be approximation-induced, and counterexamples can be semantically spurious.
\Caesar's UI/UX is designed to make these issues explicit, guide users away from unsound reasoning, and shorten the verification feedback loop.
Features are explained in detail in the extensive documentation\footnote{\url{https://www.caesarverifier.org/docs/}}.

\subsection{Installation and Command-Line Interface}
\Caesar is distributed as a single binary for Windows, Mac, and Linux.
This makes installation and usage easy, as no dependencies need to be installed separately.
The command-line interface contains many options to customize verification runs.
In particular, users can configure the backends extensively, e.g., adjusting different SMT encoding aspects.
These options and flags along with their chosen defaults make design choices explicit to users, while still allowing for experimentation when new features are added.

\subsection{Interactive Verification Environment}\label{sec:vscode-ext}
\begin{figure}[t]
    \centering
    \begin{subfigure}[t]{0.48\textwidth}
        \centering
        \includegraphics[width=\textwidth, alt={Screenshot of VSCode showing a HeyVL program of the geometric loop with post 'c' and a failing verification. The while loop is annotated with an invariant '@invariant(c + 1)' which is highlighted with an error: 'invariant might not be inductive'.}]{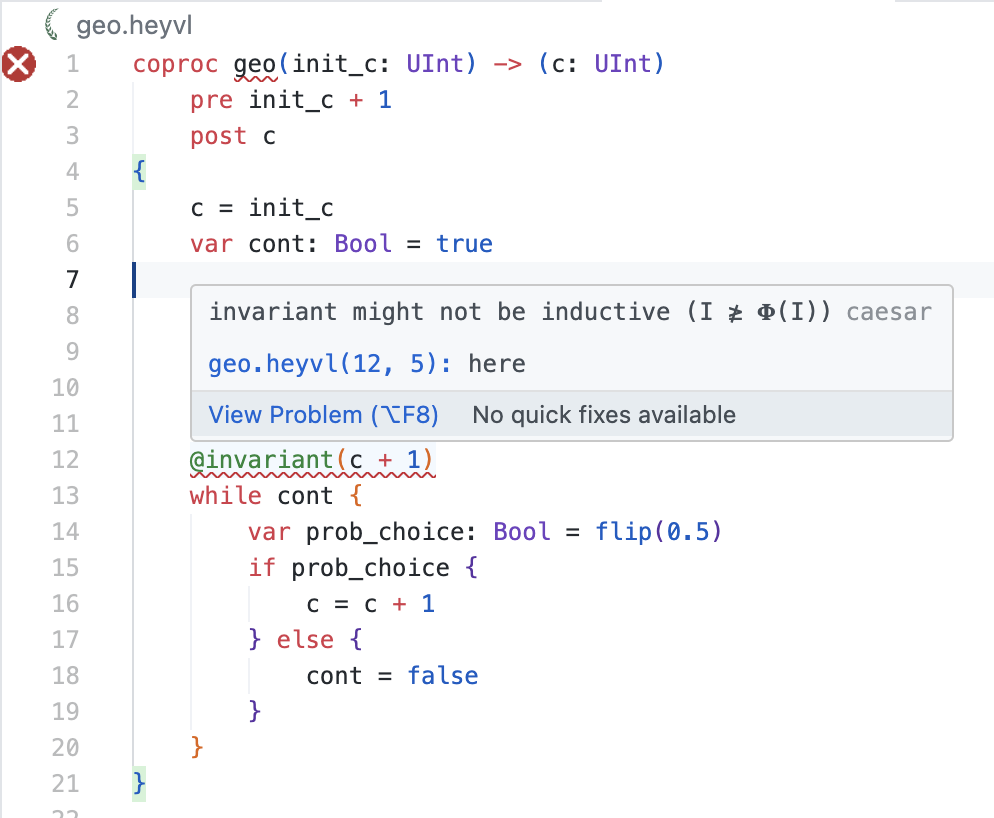}
        \caption{Identifying a non-inductive invariant.}
        \label{fig:slicing-demo}
    \end{subfigure}
    \hfill
    \begin{subfigure}[t]{0.48\textwidth}
        \centering
        \includegraphics[width=\textwidth, alt={Screenshot of VSCode showing a HeyVL program of the geometric loop with post 'c' and a successful verification. Empty lines between program statements are filled with inline verification conditions (marked with ▷) computed up to the current line.}]{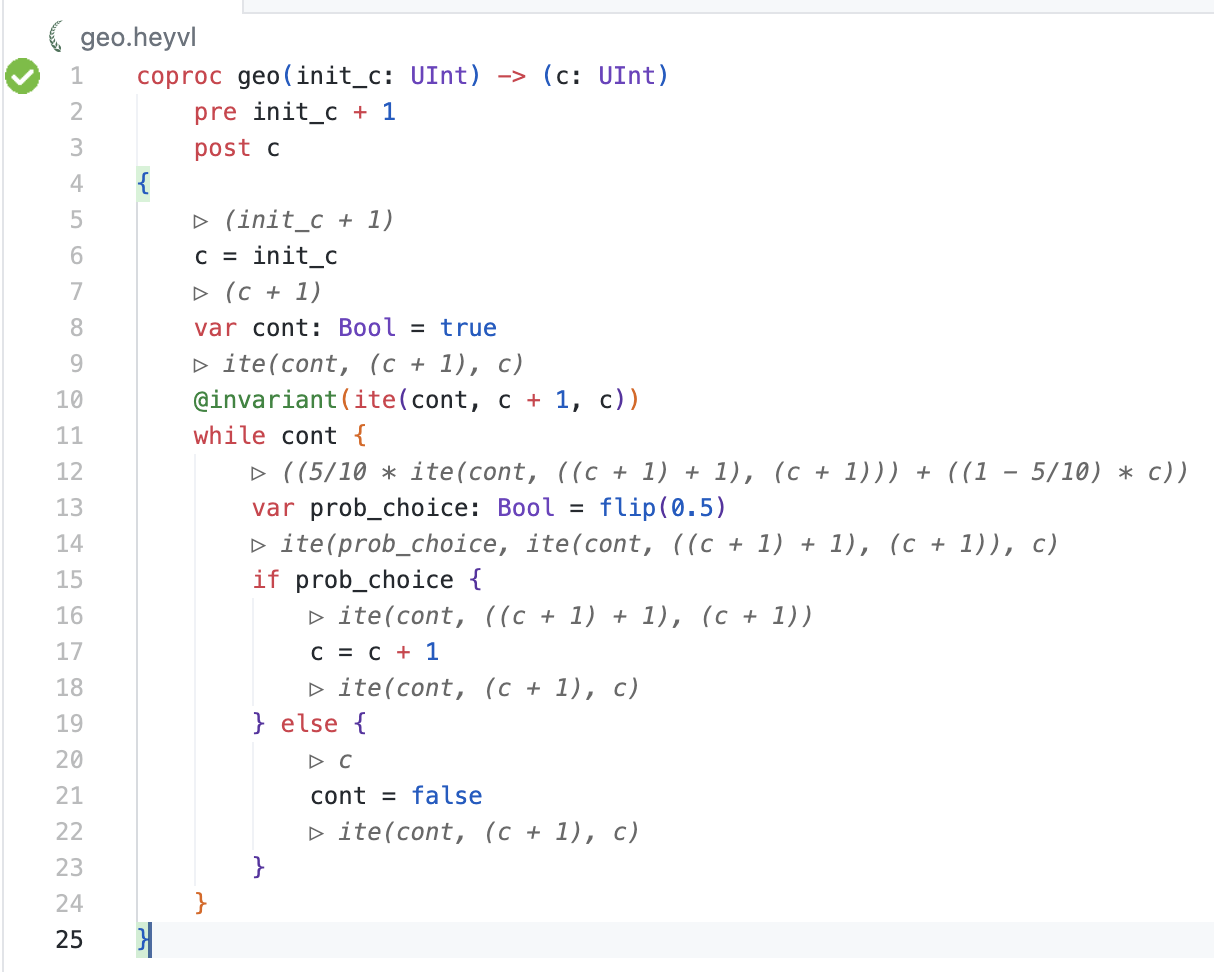}
        \caption{Inline verification conditions.}
        \label{fig:vc-demo}
    \end{subfigure}
    \caption{\Caesar's VSCode extension providing diagnostics and explanations.}
    \label{fig:diagnostics-demo}
\end{figure}

\Caesar comes with a dedicated extension for Visual Studio Code (VSCode) based on the Language Server Protocol (LSP).
This extension provides syntax highlighting, code snippets, and the ability to verify \HeyVL{} files directly from the editor on file save or on command.
It also supports automatic installation and updates of the \Caesar binary.
Verification results are shown in the gutter line and inline in the editor (\Cref{fig:slicing-demo}), and with diagnostics such as errors and warnings displayed in the code and in the "Problems" menu in VSCode.
The extension also provides inline explanations of computed verification conditions (\Cref{fig:vc-demo}), helping users understand the verification process.

\subsection{Semantic Guardrails}\label{sec:features-guardrails}

While Caesar allows arbitrary combinations of procedure kinds, original semantics, and proof rules, not all combinations are sound (\Cref{sec:proof-rules}).
We implemented diagnostics called \emph{semantic guardrails} that help users avoid unsound combinations, and give diagnostics about guaranteed soundness of proofs and refutations.

\subsection{Diagnostics: Localization and Hints}\label{sec:features-diagnostics}

\Caesar tracks source-level provenance throughout the verification pipeline, allowing failed verification conditions in \HeyLo{} to be mapped back to the originating \HeyVL{} statements.
In the probabilistic setting, a single counterexample may require multiple assertions along different execution paths.
Due to the QE, we often obtain sound counterexamples from Z3 for the verification condition.
Therefore, \Caesar soundly localizes sets of contributing verification statements.

A diagnostic then might indicate that an invariant is not inductive (seen in \Cref{fig:slicing-demo}) or that a pre is too weak.
In addition, \Caesar can give hints for verifying programs, e.g., that a loop can be replaced with a simple $\symIf$ statement.
This might alert the user to potential issues such as dead code or that the specification is too weak to require a loop.
Diagnostics are based on computing \emph{slices} of a \HeyVL{} program~\cite{brutus} that either preserve verification or refutation of the original program.

\subsection{User Experiences and Evaluation}

In addition to the applications in academic research mentioned above, \Caesar was used at the \emph{Summer School on Formal Techniques} in 2025 with $\approx 60$ students, as well as two iterations of a \emph{Probabilistic Programming} lecture with $\approx 80$ students each in the years 2024 and 2025.
Students informally reported that \Caesar's UI was helpful for learning probabilistic verification concepts and that the diagnostics helped them understand verification failures.

\Caesar was successfully applied to a number of case studies, showing its usefulness and versatility as a probabilistic verification platform.
For Rabin's mutual exclusion protocol~\cite{DBLP:conf/podc/KushilevitzR92}, \Caesar can refute bounds on success probabilities via bounded model checking, or prove exact success probabilities through a combination of proof rules~\cite{DBLP:journals/pacmpl/SchroerBKKM23}.
On a parametrized version of the bounded retransmission protocol~\cite{DBLP:conf/tacas/DArgenioKRT97}, symbolic bounds on quantitative properties were shown in~\cite{Jaarsveld2025}.
Almost-sure termination of random walks was shown with the built-in proof rule of~\cite{DBLP:journals/pacmpl/McIverMKK18}, while bounds on their expected runtime are obtained using an encoding of the expected runtime calculus $\symErt$~\cite{DBLP:journals/jacm/KaminskiKMO18}.
Bounds on the expected value of a variable sampled from the Irwin-Hall distribution are shown in~\cite{DBLP:journals/pacmpl/BatzKRW25}.
The almost-sure termination of, e.g., Ben-Or's distributed consensus protocol under weakly-fair scheduling was shown with \Caesar in~\cite{10.1145/3776691} by encoding the application of their new proof rule.

A quantitative evaluation was given in~\cite{DBLP:journals/pacmpl/SchroerBKKM23}, comparing different programs and features and showing, e.g., runtimes of \Caesar.
An evaluation of the slicing component was done in~\cite{brutus}, reporting on the overhead introduced by slicing to obtain diagnostics.
On most of the considered test cases, verification is almost instant, indicating that \Caesar is suitable for probabilistic program verification in an interactive setting.
However, since the complexity of the verification problem highly depends on the program, specification, and proof rule at hand, performance of deductive verification is highly problem-specific and thus difficult to usefully compare in a comprehensive way.
Due to the lack of a stable benchmark suite for deductive probabilistic verification, we therefore do not include another empirical evaluation.
Establishing a suitable benchmark set is an important step for future work.

\section{Related Work}\label{sec:related-work}

\paragraph{Deductive Verification.}
Classical IVLs like \toolname{Boogie}~\cite{leino2008boogie}, \toolname{Why3}~\cite{DBLP:conf/esop/FilliatreP13}, or \toolname{Viper}~\cite{viper} focus on verification of lower bounds on Boolean predicates.
They are the foundation for various verifiers, such as \toolname{Dafny}~\cite{leinoDafnyAutomaticProgram2010}, \toolname{Frama-C}~\cite{DBLP:journals/fac/KirchnerKPSY15}, or \toolname{Prusti}~\cite{AstrauskasMuellerPoliSummers19b}, which reason about partial correctness of loops with invariants and allow termination proofs through variants.
In comparison, \Caesar has a variety of built-in proof rules for probabilistic loop analysis, supporting expectation-based reasoning about upper and lower bounds.

\toolname{Dafny-VMC}~\cite{Zetzsche_Dafny-VMC_a_Library_2023} adds distribution-based reasoning about probabilistic samplers to \toolname{Dafny} by encoding probabilistic computations as transformations of random bitstreams.
Building on \toolname{Why3}, \toolname{StatWhy}~\cite{DBLP:conf/cav/KawamotoKS25} verifies programs that perform statistical hypothesis testing using a program logic with modal operators to represent beliefs.
In contrast, \Caesar uses a quantitative logic that natively allows reasoning about expected values.
As stated in~\cite{DBLP:conf/esop/BartheEGGHS18}, a detailed comparison of these approaches is difficult.

\paragraph{Probabilistic Program Verification.}
Probabilistic model checkers like \toolname{Storm}~\cite{DBLP:journals/sttt/HenselJKQV22} and \toolname{PRISM}~\cite{DBLP:conf/cav/KwiatkowskaNP11} provide automated verification for finite state probabilistic models through exhaustive state-space exploration, whereas \Caesar's deductive approach can provide symbolic guarantees on unbounded domains.

Different expectation-based verification approaches~\cite{kozenProbabilisticPDL1983,kozenProbabilisticPDL1985,mciverAbstractionRefinementProof2005,kaminskiWeakestPreconditionReasoning2016} have been formalized by~\cite{ProbabilisticGuardedCommands2005,hoelzlert} in \toolname{Isabelle/HOL}~\cite{isbabelle}.
Their focus lies on individual formalizations of the calculi's meta theory, while \Caesar aims to provide a unifying infrastructure for many verification techniques.

The theorem prover \toolname{EasyCrypt}~\cite{DBLP:conf/fosad/BartheDGKSS13} is tailored to the verification of cryptographic algorithms.
It has been extended with several program logics, including an assertion-based logic \toolname{Ellora}~\cite{DBLP:conf/esop/BartheEGGHS18} with predicates over (sub)distributions, and an expectation-based logic \toolname{eHL}~\cite{DBLP:journals/pacmpl/AvanziniBGMV24} for upper bounds.
They do not support recursion, conditioning and general non-determinism.
Existing formalizations of probabilistic program semantics by~\cite{10.1145/3732291} in the theorem prover \toolname{Rocq}, and by~\cite{Tristan_SampCert_Verified_2024} in \toolname{Lean} also do not support non-determinism or unbounded iteration.

There is a wide range of automatic analysis tools which focus on certain program properties, e.g., inferring bounds on (moments of) final distributions~\cite{DBLP:conf/cav/GehrMV16,DBLP:journals/pacmpl/MoosbruggerSBK22,DBLP:journals/pacmpl/ZaiserMO25,DBLP:conf/atva/HuangDM21,DBLP:conf/esop/HaaseBGKKKW26}, bounding expected runtimes~\cite{koat,ecoimp,leutgebAutomatedExpectedAmortised2022,ngoBoundedExpectationsResource2018} or proving (positive) almost-sure termination~\cite{amber,automatedterm,DBLP:journals/pacmpl/MajumdarS25,cegisupast}.
These methods often work on restricted language fragments.
An integration with \Caesar, analogous to the \toolname{Storm} backend, could be useful to reduce the manual verification effort (where applicable).
Similarly, synthesis approaches like~\cite{DBLP:journals/pacmpl/AvanziniMS23,batzProbabilisticProgramVerification2022,DBLP:conf/cav/AndriushchenkoC21,invsys1,invsys2,invsys3,invsys4,moments1,batzPrIC3PropertyDirected2020,rankingsuper} that automatically find, e.g., invariants, can be used to instantiate proof rules and verify programs with \Caesar.

\section{Conclusion and Future Work}

\Caesar is a deductive verifier for probabilistic programs based on the intermediate verification language \HeyVL.
\HeyVL supports reasoning about both lower and upper bounds on expected values, and through quantitative verification statements enables users to encode a wide range of proof rules.
The verification condition generator translates \HeyVL programs into verification conditions in the \HeyLo logic, which are then checked by the SMT solver Z3.
\Caesar also includes a probabilistic model checking backend that translates a subset of \HeyVL programs into MDPs for expected value computation via probabilistic model checking.
\Caesar includes several usability features, including diagnostics for error localization, hints, and to prevent unsound reasoning.

In future work, we plan to incorporate invariant synthesis approaches into \Caesar to ease the verification process.
This includes classical approaches described in~\Cref{sec:related-work} and LLM-assisted invariant discovery.
Transpiling code from other probabilistic programming languages for verification with \Caesar, or transpiling verified \HeyVL code to obtain executable, correct programs, are further efforts that will expand the tool's applicability and allow usage within existing software projects.
To apply \Caesar in larger software projects (in particular in machine learning), we foresee challenges due to the heavy reliance on sampling from continuous distributions.
Therefore, we plan to develop an extended standard library of (continuous) distributions and data structures.
Finally, establishing a standard benchmark suite for deductive probabilistic program verification will allow for comparisons across tools and techniques.

\ifAnon{}{
	\section*{Data-Availability Statement}
	The tool \Caesar, its VSCode extension, as well as benchmark programs that were submitted for the artifact evaluation are publicly available on Zenodo~\cite{artifact}.
	We also develop our tools as open-source software at \url{https://github.com/moves-rwth/caesar}.

	\section*{Acknowledgments}
	This work was partially supported by the ERC POC Grant VERIPROB (grant no.\ 101158076), the Research Training Group 2236 \emph{UnRAVeL} (project number 282652900), DFF-1 project AuRoRa, and by the European Union's Horizon 2020 research and innovation programme under the Marie Skłodowska-Curie grant agreement MISSION (grant no.\ 101008233).
	We thank Emil Beothy-Elo, Franka van Jaarsveld, and Se Rin Yang for their contributions to \Caesar.
	We thank Alexander Bork and Tobias Winkler for their feedback on an early draft.

	\marknewtext{
		\subsubsection*{Disclosure of Interests.}
		The authors have no competing interests to declare that are relevant to the content of this article.
	}

}

\printbibliography

@article{DBLP:journals/pacmpl/SchroerBKKM23,
  author       = {Philipp Schr{\"{o}}er and
                  Kevin Batz and
                  Benjamin Lucien Kaminski and
                  Joost{-}Pieter Katoen and
                  Christoph Matheja},
  title        = {A Deductive Verification Infrastructure for Probabilistic Programs},
  journal      = {Proc. {ACM} Program. Lang.},
  volume       = {7},
  number       = {{OOPSLA2}},
  pages        = {2052--2082},
  year         = {2023},
  url          = {https://doi.org/10.1145/3622870},
  doi          = {10.1145/3622870},
  bibsource    = {dblp computer science bibliography, https://dblp.org}
}

@inproceedings{DBLP:conf/esop/FilliatreP13,
  author       = {Jean{-}Christophe Filli{\^{a}}tre and
                  Andrei Paskevich},
  title        = {Why3 - Where Programs Meet Provers},
  booktitle    = {{ESOP}},
  series       = {Lecture Notes in Computer Science},
  volume       = {7792},
  pages        = {125--128},
  publisher    = {Springer},
  year         = {2013},
  doi = {10.1007/978-3-642-37036-6_8}
}

@online{leino2008boogie,
author = {Leino, K. Rustan M.},
title = {This is Boogie 2},
year = {2008},
month = {06},
url = {https://www.microsoft.com/en-us/research/publication/this-is-boogie-2-2/},
}

@inproceedings{viper,
  author       = {Peter M{\"{u}}ller and
                  Malte Schwerhoff and
                  Alexander J. Summers},
  editor       = {Barbara Jobstmann and
                  K. Rustan M. Leino},
  title        = {Viper: {A} Verification Infrastructure for Permission-Based Reasoning},
  booktitle    = {Verification, Model Checking, and Abstract Interpretation - 17th International
                  Conference, {VMCAI} 2016, St. Petersburg, FL, USA, January 17-19,
                  2016. Proceedings},
  series       = {Lecture Notes in Computer Science},
  volume       = {9583},
  pages        = {41--62},
  publisher    = {Springer},
  year         = {2016},
  url          = {https://doi.org/10.1007/978-3-662-49122-5\_2},
  doi          = {10.1007/978-3-662-49122-5\_2},
  bibsource    = {dblp computer science bibliography, https://dblp.org}
}

@online{BarFT-SMTLIB,
  author =	 {Clark Barrett and Pascal Fontaine and Cesare Tinelli},
  title =	 {{The Satisfiability Modulo Theories Library (SMT-LIB)}},
  url = {https://smt-lib.org},
  year =	 2016,
}

@inproceedings{demouraZ3EfficientSMT2008,
  title = {Z3: {{An Efficient SMT Solver}}},
  shorttitle = {Z3},
  booktitle = {Tools and {{Algorithms}} for the {{Construction}} and {{Analysis}} of {{Systems}}},
  author = {{de Moura}, Leonardo and Bj{\o}rner, Nikolaj},
  editor = {Ramakrishnan, C. R. and Rehof, Jakob},
  year = {2008},
  series = {Lecture {{Notes}} in {{Computer Science}}},
  publisher = {{Springer}},
  address = {{Berlin, Heidelberg}},
  isbn = {978-3-540-78800-3},
  doi = {10.1007/978-3-540-78800-3\_24}
}

@inproceedings{DBLP:conf/tacas/BuddeDHHJT17,
  author       = {Carlos E. Budde and
                  Christian Dehnert and
                  Ernst Moritz Hahn and
                  Arnd Hartmanns and
                  Sebastian Junges and
                  Andrea Turrini},
  editor       = {Axel Legay and
                  Tiziana Margaria},
  title        = {{JANI:} Quantitative Model and Tool Interaction},
  booktitle    = {Tools and Algorithms for the Construction and Analysis of Systems
                  - 23rd International Conference, {TACAS} 2017, Held as Part of the
                  European Joint Conferences on Theory and Practice of Software, {ETAPS}
                  2017, Uppsala, Sweden, April 22-29, 2017, Proceedings, Part {II}},
  series       = {Lecture Notes in Computer Science},
  volume       = {10206},
  pages        = {151--168},
  year         = {2017},
  url          = {https://doi.org/10.1007/978-3-662-54580-5\_9},
  doi          = {10.1007/978-3-662-54580-5\_9},
  bibsource    = {dblp computer science bibliography, https://dblp.org}
}

@article{DBLP:journals/sttt/HenselJKQV22,
  author       = {Christian Hensel and
                  Sebastian Junges and
                  Joost{-}Pieter Katoen and
                  Tim Quatmann and
                  Matthias Volk},
  title        = {The probabilistic model checker Storm},
  journal      = {Int. J. Softw. Tools Technol. Transf.},
  volume       = {24},
  number       = {4},
  pages        = {589--610},
  year         = {2022},
  url          = {https://doi.org/10.1007/s10009-021-00633-z},
  doi          = {10.1007/S10009-021-00633-Z},
  bibsource    = {dblp computer science bibliography, https://dblp.org}
}

@article{DBLP:journals/pacmpl/0001BHKKM21,
  author       = {Alejandro Aguirre and
                  Gilles Barthe and
                  Justin Hsu and
                  Benjamin Lucien Kaminski and
                  Joost{-}Pieter Katoen and
                  Christoph Matheja},
  title        = {A pre-expectation calculus for probabilistic sensitivity},
  journal      = {Proc. {ACM} Program. Lang.},
  volume       = {5},
  number       = {{POPL}},
  pages        = {1--28},
  year         = {2021},
  url          = {https://doi.org/10.1145/3434333},
  doi          = {10.1145/3434333},
  bibsource    = {dblp computer science bibliography, https://dblp.org}
}

@article{DBLP:journals/pacmpl/BatzKRW25,
  author       = {Kevin Batz and
                  Joost{-}Pieter Katoen and
                  Francesca Randone and
                  Tobias Winkler},
  title        = {Foundations for Deductive Verification of Continuous Probabilistic
                  Programs: From Lebesgue to Riemann and Back},
  journal      = {Proc. {ACM} Program. Lang.},
  volume       = {9},
  number       = {{OOPSLA1}},
  pages        = {421--448},
  year         = {2025},
  url          = {https://doi.org/10.1145/3720429},
  doi          = {10.1145/3720429},
  bibsource    = {dblp computer science bibliography, https://dblp.org}
}

@inproceedings{DBLP:conf/vecos/Matheja24,
  author       = {Christoph Matheja},
  editor       = {Bernhard Steffen},
  title        = {A Game-Based Semantics for the Probabilistic Intermediate Verification
                  Language HeyVL},
  booktitle    = {Bridging the Gap Between {AI} and Reality - Second International Conference,
                  AISoLA 2024, Crete, Greece, October 30 - November 3, 2024, Proceedings},
  series       = {Lecture Notes in Computer Science},
  volume       = {15217},
  pages        = {242--258},
  publisher    = {Springer},
  year         = {2024},
  url          = {https://doi.org/10.1007/978-3-031-75434-0\_17},
  doi          = {10.1007/978-3-031-75434-0\_17},
  bibsource    = {dblp computer science bibliography, https://dblp.org}
}

@article{10.1145/3776691,
author = {Enea, Constantin and Majumdar, Rupak and Motwani, Harshit Jitendra and Sathiyanarayana, V. R.},
title = {Verifying Almost-Sure Termination for Randomized Distributed Algorithms},
year = {2026},
issue_date = {January 2026},
publisher = {Association for Computing Machinery},
address = {New York, NY, USA},
volume = {10},
number = {POPL},
url = {https://doi.org/10.1145/3776691},
doi = {10.1145/3776691},
journal = {Proc. ACM Program. Lang.},
month = jan,
articleno = {49},
numpages = {30}
}

@book{mciverAbstractionRefinementProof2005,
  title = {Abstraction, {{Refinement}} and {{Proof}} for {{Probabilistic Systems}}},
  author = {McIver, Annabelle and Morgan, Charles Carroll},
  year = {2005},
  series = {Monographs in {{Computer Science}}},
  publisher = {{Springer-Verlag}},
  address = {{New York}},
  isbn = {978-0-387-40115-7},
  doi = {10.1007/b138392}
}

@phdthesis{kaminskiAdvancedWeakestPrecondition2019,
  title = {Advanced {{Weakest Precondition Calculi}} for {{Probabilistic Programs}}},
  author = {Kaminski, Benjamin Lucien},
  year = {2019},
  month = feb,
  school = {RWTH Aachen University},
  doi = {10.18154/RWTH-2019-01829}
}

@article{batzRelativelyCompleteVerification2021,
	author       = {Kevin Batz and
	Benjamin Lucien Kaminski and
	Joost{-}Pieter Katoen and
	Christoph Matheja},
	title        = {Relatively complete verification of probabilistic programs: an expressive
	language for expectation-based reasoning},
	journal      = {Proc. {ACM} Program. Lang.},
	volume       = {5},
	number       = {{POPL}},
	pages        = {1--30},
	year         = {2021},
	doi = {10.1145/3434320}
}

@MASTERSTHESIS{Schroer:998370,
      author       = {Schroer, Philipp},
      othercontributors = {Katoen, Joost-Pieter and Müller, Peter and Batz, Kevin and
                          Kaminski, Benjamin and Matheja, Christoph},
      title        = {{A} deductive verifier for probabilistic programs},
      school       = {RWTH Aachen University},
      address      = {Aachen},
      publisher    = {RWTH Aachen University},
      reportid     = {RWTH-2024-11340},
      pages        = {1 Online-Ressource: Illustrationen},
      year         = {2024},
      note         = {Veröffentlicht auf dem Publikationsserver der RWTH Aachen
                      University 2024; Masterarbeit, RWTH Aachen University, 2022},
      cin          = {121310 / 120000},
      ddc          = {004},
      cid          = {$I:(DE-82)121310_20140620$ / $I:(DE-82)120000_20140620$},
      typ          = {PUB:(DE-HGF)19},
      doi          = {10.18154/RWTH-2024-11340},
      url          = {https://publications.rwth-aachen.de/record/998370},
}

@article{parkFixpointInductionProofs1969,
  title = {Fixpoint Induction and Proofs of Program Properties},
  author = {Park, David},
  year = {1969},
  journal = {Machine Intelligence},
  volume = {5},
  publisher = {{Edinburgh University Press}}
}

@inproceedings{DBLP:conf/cav/BatzCKKMS20,
  author       = {Kevin Batz and
                  Mingshuai Chen and
                  Benjamin Lucien Kaminski and
                  Joost{-}Pieter Katoen and
                  Christoph Matheja and
                  Philipp Schr{\"{o}}er},
  editor       = {Alexandra Silva and
                  K. Rustan M. Leino},
  title        = {Latticed k-Induction with an Application to Probabilistic Programs},
  booktitle    = {Computer Aided Verification - 33rd International Conference, {CAV}
                  2021, Virtual Event, July 20-23, 2021, Proceedings, Part {II}},
  series       = {Lecture Notes in Computer Science},
  volume       = {12760},
  pages        = {524--549},
  publisher    = {Springer},
  year         = {2021},
  url          = {https://doi.org/10.1007/978-3-030-81688-9\_25},
  doi          = {10.1007/978-3-030-81688-9\_25},
  bibsource    = {dblp computer science bibliography, https://dblp.org}
}

@article{DBLP:journals/pacmpl/HarkKGK20,
  author       = {Marcel Hark and
                  Benjamin Lucien Kaminski and
                  J{\"{u}}rgen Giesl and
                  Joost{-}Pieter Katoen},
  title        = {Aiming low is harder: induction for lower bounds in probabilistic
                  program verification},
  journal      = {Proc. {ACM} Program. Lang.},
  volume       = {4},
  number       = {{POPL}},
  pages        = {37:1--37:28},
  year         = {2020},
  url          = {https://doi.org/10.1145/3371105},
  doi          = {10.1145/3371105},
  bibsource    = {dblp computer science bibliography, https://dblp.org}
}

@article{DBLP:journals/pacmpl/McIverMKK18,
  author       = {Annabelle McIver and
                  Carroll Morgan and
                  Benjamin Lucien Kaminski and
                  Joost{-}Pieter Katoen},
  title        = {A new proof rule for almost-sure termination},
  journal      = {Proc. {ACM} Program. Lang.},
  volume       = {2},
  number       = {{POPL}},
  pages        = {33:1--33:28},
  year         = {2018},
  url          = {https://doi.org/10.1145/3158121},
  doi          = {10.1145/3158121},
  bibsource    = {dblp computer science bibliography, https://dblp.org}
}

@inproceedings{DBLP:conf/cav/ChakarovS13,
  author    = {Aleksandar Chakarov and
               Sriram Sankaranarayanan},
  editor    = {Natasha Sharygina and
               Helmut Veith},
  title     = {Probabilistic Program Analysis with Martingales},
  booktitle = {Computer Aided Verification - 25th International Conference, {CAV}
               2013, Saint Petersburg, Russia, July 13-19, 2013. Proceedings},
  series    = {Lecture Notes in Computer Science},
  volume    = {8044},
  pages     = {511--526},
  publisher = {Springer},
  year      = {2013},
  url       = {https://doi.org/10.1007/978-3-642-39799-8\_34},
  doi       = {10.1007/978-3-642-39799-8\_34},
  bibsource = {dblp computer science bibliography, https://dblp.org}
}

@inproceedings{DBLP:conf/popl/FioritiH15,
  author       = {Luis Mar{\'{\i}}a Ferrer Fioriti and
                  Holger Hermanns},
  editor       = {Sriram K. Rajamani and
                  David Walker},
  title        = {Probabilistic Termination: Soundness, Completeness, and Compositionality},
  booktitle    = {Proceedings of the 42nd Annual {ACM} {SIGPLAN-SIGACT} Symposium on
                  Principles of Programming Languages, {POPL} 2015, Mumbai, India, January
                  15-17, 2015},
  pages        = {489--501},
  publisher    = {{ACM}},
  year         = {2015},
  url          = {https://doi.org/10.1145/2676726.2677001},
  doi          = {10.1145/2676726.2677001},
  bibsource    = {dblp computer science bibliography, https://dblp.org}
}

@article{DBLP:journals/acta/KaminskiKM19,
  author       = {Benjamin Lucien Kaminski and
                  Joost{-}Pieter Katoen and
                  Christoph Matheja},
  title        = {On the hardness of analyzing probabilistic programs},
  journal      = {Acta Informatica},
  volume       = {56},
  number       = {3},
  pages        = {255--285},
  year         = {2019},
  url          = {https://doi.org/10.1007/s00236-018-0321-1},
  doi          = {10.1007/S00236-018-0321-1},
  bibsource    = {dblp computer science bibliography, https://dblp.org}
}

@phdthesis{DBLP:phd/ethos/Jones90,
  author       = {Claire Jones},
  title        = {Probabilistic non-determinism},
  school       = {University of Edinburgh, {UK}},
  year         = {1990},
  url          = {https://hdl.handle.net/1842/413},
  bibsource    = {dblp computer science bibliography, https://dblp.org}
}

@incollection{DBLP:books/mc/22/Hahnle22,
  author       = {Reiner H{\"{a}}hnle},
  editor       = {Krzysztof R. Apt and
                  Tony Hoare},
  title        = {Dijkstra's Legacy on Program Verification},
  booktitle    = {Edsger Wybe Dijkstra: His Life, Work, and Legacy},
  series       = {{ACM} Books},
  volume       = {45},
  pages        = {105--140},
  publisher    = {{ACM} / Morgan {\&} Claypool},
  year         = {2022},
  url          = {https://doi.org/10.1145/3544585.3544593},
  doi          = {10.1145/3544585.3544593},
  bibsource    = {dblp computer science bibliography, https://dblp.org}
}

@book{DBLP:series/lncs/10001,
  editor       = {Wolfgang Ahrendt and
                  Bernhard Beckert and
                  Richard Bubel and
                  Reiner H{\"{a}}hnle and
                  Peter H. Schmitt and
                  Mattias Ulbrich},
  title        = {Deductive Software Verification - The KeY Book - From Theory to Practice},
  series       = {Lecture Notes in Computer Science},
  volume       = {10001},
  publisher    = {Springer},
  year         = {2016},
  url          = {https://doi.org/10.1007/978-3-319-49812-6},
  doi          = {10.1007/978-3-319-49812-6},
  isbn         = {978-3-319-49811-9},
  bibsource    = {dblp computer science bibliography, https://dblp.org}
}

@MASTERSTHESIS{BeothyElo:1016969,
      author       = {Beothy-Elo, Emil},
      othercontributors = {Katoen, Joost-Pieter and Noll, Thomas and Schroer, Philipp
                          and Haase, Darion},
      title        = {{E}ffective quantifier-based reasoning for quantitative
                      deductive verification},
      school       = {RWTH Aachen University},
      address      = {Aachen},
      publisher    = {RWTH Aachen University},
      reportid     = {RWTH-2025-07050},
      year         = {2025},
      note         = {Veröffentlicht auf dem Publikationsserver der RWTH Aachen
                      University; Masterarbeit, RWTH Aachen University, 2025},
      cin          = {121310 / 120000},
      ddc          = {004},
      cid          = {$I:(DE-82)121310_20140620$ / $I:(DE-82)120000_20140620$},
      typ          = {PUB:(DE-HGF)19},
      doi          = {10.18154/RWTH-2025-07050},
      url          = {https://publications.rwth-aachen.de/record/1016969},
}

@inproceedings{brutus,
  author       = {Philipp Schr{\"{o}}er and
                  Darion Haase and
                  Joost{-}Pieter Katoen},
  editor       = {Robbert Krebbers},
  title        = {Error Localization, Certificates, and Hints for Probabilistic Program
                  Verification via Slicing},
  booktitle    = {Programming Languages and Systems - 35th European Symposium on Programming,
                  {ESOP} 2026, Held as Part of the International Joint Conferences on
                  Theory and Practice of Software, {ETAPS} 2026, Turin, Italy, April
                  11-16, 2026, Proceedings, Part {II}},
  series       = {Lecture Notes in Computer Science},
  pages        = {210--240},
  publisher    = {Springer},
  year         = {2026},
  url          = {https://doi.org/10.1007/978-3-032-22723-2\_8},
  doi          = {10.1007/978-3-032-22723-2\_8},
  bibsource    = {dblp computer science bibliography, https://dblp.org}
}

@inproceedings{DBLP:conf/cpaior/LiffitonM13,
  author       = {Mark H. Liffiton and
                  Ammar Malik},
  editor       = {Carla P. Gomes and
                  Meinolf Sellmann},
  title        = {Enumerating Infeasibility: Finding Multiple MUSes Quickly},
  booktitle    = {Integration of {AI} and {OR} Techniques in Constraint Programming
                  for Combinatorial Optimization Problems, 10th International Conference,
                  {CPAIOR} 2013, Yorktown Heights, NY, USA, May 18-22, 2013. Proceedings},
  series       = {Lecture Notes in Computer Science},
  volume       = {7874},
  pages        = {160--175},
  publisher    = {Springer},
  year         = {2013},
  url          = {https://doi.org/10.1007/978-3-642-38171-3\_11},
  doi          = {10.1007/978-3-642-38171-3\_11},
  bibsource    = {dblp computer science bibliography, https://dblp.org}
}

@article{DBLP:journals/theoretics/BrazdilCCFKKM0U25,
  author       = {Tom{\'{a}}s Br{\'{a}}zdil and
                  Krishnendu Chatterjee and
                  Martin Chmelik and
                  Vojtech Forejt and
                  Jan Kret{\'{\i}}nsk{\'{y}} and
                  Marta Kwiatkowska and
                  Tobias Meggendorfer and
                  David Parker and
                  Mateusz Ujma},
  title        = {Learning Algorithms for Verification of Markov Decision Processes},
  journal      = {TheoretiCS},
  volume       = {4},
  year         = {2025},
  url          = {https://doi.org/10.46298/theoretics.25.10},
  doi          = {10.46298/THEORETICS.25.10},
  bibsource    = {dblp computer science bibliography, https://dblp.org}
}

@inproceedings{leinoDafnyAutomaticProgram2010,
  title = {Dafny: {{An Automatic Program Verifier}} for {{Functional Correctness}}},
  shorttitle = {Dafny},
  booktitle = {Logic for {{Programming}}, {{Artificial Intelligence}}, and {{Reasoning}}},
  author = {Leino, K. Rustan M.},
  editor = {Clarke, Edmund M. and Voronkov, Andrei},
  year = {2010},
  series = {Lecture {{Notes}} in {{Computer Science}}},
  publisher = {{Springer}},
  address = {{Berlin, Heidelberg}},
  isbn = {978-3-642-17511-4},
  doi = {10.1007/978-3-642-17511-4_20}
}

@article{DBLP:journals/toplas/OlmedoGJKKM18,
  author       = {Federico Olmedo and
                  Friedrich Gretz and
                  Nils Jansen and
                  Benjamin Lucien Kaminski and
                  Joost{-}Pieter Katoen and
                  Annabelle McIver},
  title        = {Conditioning in Probabilistic Programming},
  journal      = {{ACM} Trans. Program. Lang. Syst.},
  volume       = {40},
  number       = {1},
  pages        = {4:1--4:50},
  year         = {2018},
  url          = {https://doi.org/10.1145/3156018},
  doi          = {10.1145/3156018},
  bibsource    = {dblp computer science bibliography, https://dblp.org}
}

@article{DBLP:journals/jacm/KaminskiKMO18,
  author    = {Benjamin Lucien Kaminski and
               Joost{-}Pieter Katoen and
               Christoph Matheja and
               Federico Olmedo},
  title     = {Weakest Precondition Reasoning for Expected Runtimes of Randomized
               Algorithms},
  journal   = {J. {ACM}},
  volume    = {65},
  number    = {5},
  pages     = {30:1--30:68},
  year      = {2018},
  url       = {https://doi.org/10.1145/3208102},
  doi       = {10.1145/3208102},
  bibsource = {dblp computer science bibliography, https://dblp.org}
}

@inproceedings{DBLP:conf/podc/KushilevitzR92,
  author       = {Eyal Kushilevitz and
                  Michael O. Rabin},
  editor       = {Norman C. Hutchinson},
  title        = {Randomized Mutual Exclusion Algorithms Revisited},
  booktitle    = {Proceedings of the Eleventh Annual {ACM} Symposium on Principles of
                  Distributed Computing, Vancouver, British Columbia, Canada, August
                  10-12, 1992},
  pages        = {275--283},
  publisher    = {{ACM}},
  year         = {1992},
  url          = {https://doi.org/10.1145/135419.135468},
  doi          = {10.1145/135419.135468},
  bibsource    = {dblp computer science bibliography, https://dblp.org}
}

@inproceedings{DBLP:conf/tacas/DArgenioKRT97,
  author       = {Pedro R. D'Argenio and
                  Joost{-}Pieter Katoen and
                  Theo C. Ruys and
                  Jan Tretmans},
  editor       = {Ed Brinksma},
  title        = {The Bounded Retransmission Protocol Must Be on Time!},
  booktitle    = {Tools and Algorithms for Construction and Analysis of Systems, Third
                  International Workshop, {TACAS} '97, Enschede, The Netherlands, April
                  2-4, 1997, Proceedings},
  series       = {Lecture Notes in Computer Science},
  volume       = {1217},
  pages        = {416--431},
  publisher    = {Springer},
  year         = {1997},
  url          = {https://doi.org/10.1007/BFb0035403},
  doi          = {10.1007/BFB0035403},
  bibsource    = {dblp computer science bibliography, https://dblp.org}
}

@mastersthesis{Jaarsveld2025,
  month    = {5},
  author   = {Jaarsveld, Franka van},
  year     = {2025},
  school   = {University of Twente},
  address  = {Enschede},
  type     = {Thesis},
  title    = {Practical Probabilistic Program Verification using Caesar},
  url      = { https://purl.utwente.nl/essays/106318}
}

@inproceedings{highlyIncremental,
  author       = {Philipp Schr{\"{o}}er and
                  Joost{-}Pieter Katoen},
  title        = {Highly Incremental: {A} Simple Programmatic Approach for Many Objectives},
  booktitle    = {Formal Methods - 27th International Symposium, {FM} 2026,
                  Tokyo, Japan, May 18--22, 2026, Proceedings},
  series       = {Lecture Notes in Computer Science},
  publisher    = {Springer},
  address      = {Cham},
  year         = {2026},
  note         = {To appear}
}

@article{DBLP:journals/pacmpl/FengCSKKZ23,
  author       = {Shenghua Feng and
                  Mingshuai Chen and
                  Han Su and
                  Benjamin Lucien Kaminski and
                  Joost{-}Pieter Katoen and
                  Naijun Zhan},
  title        = {Lower Bounds for Possibly Divergent Probabilistic Programs},
  journal      = {Proc. {ACM} Program. Lang.},
  volume       = {7},
  number       = {{OOPSLA1}},
  pages        = {696--726},
  year         = {2023},
  url          = {https://doi.org/10.1145/3586051},
  doi          = {10.1145/3586051},
  bibsource    = {dblp computer science bibliography, https://dblp.org}
}

@online{artifact,
  author       = {Schröer, Philipp and
                  Batz, Kevin and
                  Dural, Umut Yiğit and
                  Haase, Darion and
                  Kaminski, Benjamin Lucien and
                  Katoen, Joost-Pieter and
                  Matheja, Christoph},
  title        = {Caesar: A Deductive Verifier for Probabilistic
                   Programs (CAV26 Artifact)
                  },
  month        = apr,
  year         = 2026,
  organization = {Zenodo},
  doi          = {10.5281/zenodo.19838026},
  url          = {https://doi.org/10.5281/zenodo.19838026},
}

@online{Zetzsche_Dafny-VMC_a_Library_2023,
  author = {Zetzsche, Stefan and Tristan, Jean-Baptiste},
  title  = {{Dafny-VMC: a Library for Verified Monte Carlo Algorithms}},
  year   = {2023},
  note   = {Software},
  url    = {https://github.com/dafny-lang/Dafny-VMC},
}

@article{DBLP:journals/fac/KirchnerKPSY15,
  author       = {Florent Kirchner and
                  Nikolai Kosmatov and
                  Virgile Prevosto and
                  Julien Signoles and
                  Boris Yakobowski},
  title        = {Frama-C: {A} software analysis perspective},
  journal      = {Formal Aspects Comput.},
  volume       = {27},
  number       = {3},
  pages        = {573--609},
  year         = {2015},
  url          = {https://doi.org/10.1007/s00165-014-0326-7},
  doi          = {10.1007/S00165-014-0326-7},
  bibsource    = {dblp computer science bibliography, https://dblp.org}
}

@InProceedings{AstrauskasMuellerPoliSummers19b,
  title = {Leveraging {R}ust Types for Modular Specification and Verification},
  author = {V. Astrauskas and P. M\"uller and F. Poli and A. J. Summers},
  booktitle = {Object-Oriented Programming Systems, Languages, and Applications (OOPSLA)},
  journal = {Proc. ACM Program. Lang.},
  issue_date = {October 2019},
  volume = {3},
  number = {OOPSLA},
  year = {2019},
  pages = {147:1--147:30},
  url = {https://doi.org/10.1145/3360573},
  doi = {10.1145/3360573},
  publisher = {ACM}
}

@inproceedings{DBLP:conf/cav/KwiatkowskaNP11,
  author       = {Marta Z. Kwiatkowska and
                  Gethin Norman and
                  David Parker},
  editor       = {Ganesh Gopalakrishnan and
                  Shaz Qadeer},
  title        = {{PRISM} 4.0: Verification of Probabilistic Real-Time Systems},
  booktitle    = {Computer Aided Verification - 23rd International Conference, {CAV}
                  2011, Snowbird, UT, USA, July 14-20, 2011. Proceedings},
  series       = {Lecture Notes in Computer Science},
  pages        = {585--591},
  publisher    = {Springer},
  year         = {2011},
  url          = {https://doi.org/10.1007/978-3-642-22110-1\_47},
  doi          = {10.1007/978-3-642-22110-1\_47},
  bibsource    = {dblp computer science bibliography, https://dblp.org}
}

@inproceedings{DBLP:conf/fosad/BartheDGKSS13,
  author       = {Gilles Barthe and
                  Fran{\c{c}}ois Dupressoir and
                  Benjamin Gr{\'{e}}goire and
                  C{\'{e}}sar Kunz and
                  Benedikt Schmidt and
                  Pierre{-}Yves Strub},
  editor       = {Alessandro Aldini and
                  Javier L{\'{o}}pez and
                  Fabio Martinelli},
  title        = {EasyCrypt: {A} Tutorial},
  booktitle    = {Foundations of Security Analysis and Design {VII} - {FOSAD} 2012/2013
                  Tutorial Lectures},
  series       = {Lecture Notes in Computer Science},
  pages        = {146--166},
  publisher    = {Springer},
  year         = {2013},
  url          = {https://doi.org/10.1007/978-3-319-10082-1\_6},
  doi          = {10.1007/978-3-319-10082-1\_6},
  bibsource    = {dblp computer science bibliography, https://dblp.org}
}

@inproceedings{DBLP:conf/esop/BartheEGGHS18,
  author       = {Gilles Barthe and
                  Thomas Espitau and
                  Marco Gaboardi and
                  Benjamin Gr{\'{e}}goire and
                  Justin Hsu and
                  Pierre{-}Yves Strub},
  editor       = {Amal Ahmed},
  title        = {An Assertion-Based Program Logic for Probabilistic Programs},
  booktitle    = {Programming Languages and Systems - 27th European Symposium on Programming,
                  {ESOP} 2018, Held as Part of the European Joint Conferences on Theory
                  and Practice of Software, {ETAPS} 2018, Thessaloniki, Greece, April
                  14-20, 2018, Proceedings},
  series       = {Lecture Notes in Computer Science},
  pages        = {117--144},
  publisher    = {Springer},
  year         = {2018},
  url          = {https://doi.org/10.1007/978-3-319-89884-1\_5},
  doi          = {10.1007/978-3-319-89884-1\_5},
  bibsource    = {dblp computer science bibliography, https://dblp.org}
}

@article{DBLP:journals/pacmpl/AvanziniBGMV24,
  author       = {Martin Avanzini and
                  Gilles Barthe and
                  Benjamin Gr{\'{e}}goire and
                  Georg Moser and
                  Gabriele Vanoni},
  title        = {Hopping Proofs of Expectation-Based Properties: Applications to Skiplists
                  and Security Proofs},
  journal      = {Proc. {ACM} Program. Lang.},
  volume       = {8},
  number       = {{OOPSLA1}},
  pages        = {784--809},
  year         = {2024},
  url          = {https://doi.org/10.1145/3649839},
  doi          = {10.1145/3649839},
  bibsource    = {dblp computer science bibliography, https://dblp.org}
}

@inproceedings{DBLP:conf/cav/KawamotoKS25,
  author       = {Yusuke Kawamoto and
                  Kentaro Kobayashi and
                  Kohei Suenaga},
  editor       = {Ruzica Piskac and
                  Zvonimir Rakamaric},
  title        = {StatWhy: Formal Verification Tool for Statistical Hypothesis Testing
                  Programs},
  booktitle    = {Computer Aided Verification - 37th International Conference, {CAV}
                  2025, Zagreb, Croatia, July 23-25, 2025, Proceedings, Part {II}},
  series       = {Lecture Notes in Computer Science},
  pages        = {216--230},
  publisher    = {Springer},
  year         = {2025},
  url          = {https://doi.org/10.1007/978-3-031-98679-6\_10},
  doi          = {10.1007/978-3-031-98679-6\_10},
  bibsource    = {dblp computer science bibliography, https://dblp.org}
}

@online{Tristan_SampCert_Verified_2024,
  author = {Tristan, Jean-Baptiste},
  doi = {10.5281/zenodo.11204806},
  month = may,
  title = {{SampCert : Verified Differential Privacy}},
  url = {https://github.com/leanprover/SampCert},
  version = {1.0.0},
  year = {2024}
}

@article{10.1145/3732291,
  author = {Affeldt, Reynald and Cohen, Cyril and Saito, Ayumu},
  title = {Semantics of Probabilistic Programs Using s-Finite Kernels in Dependent Type Theory},
  year = {2025},
  issue_date = {September 2025},
  publisher = {Association for Computing Machinery},
  address = {New York, NY, USA},
  volume = {1},
  number = {3},
  url = {https://doi.org/10.1145/3732291},
  doi = {10.1145/3732291},
  journal = {ACM Trans. Probab. Mach. Learn.},
  month = aug,
  articleno = {16},
  numpages = {34}
}

@inproceedings{DBLP:conf/cav/GehrMV16,
  author       = {Timon Gehr and
                  Sasa Misailovic and
                  Martin T. Vechev},
  editor       = {Swarat Chaudhuri and
                  Azadeh Farzan},
  title        = {{PSI:} Exact Symbolic Inference for Probabilistic Programs},
  booktitle    = {Computer Aided Verification - 28th International Conference, {CAV}
                  2016, Toronto, ON, Canada, July 17-23, 2016, Proceedings, Part {I}},
  series       = {Lecture Notes in Computer Science},
  pages        = {62--83},
  publisher    = {Springer},
  year         = {2016},
  url          = {https://doi.org/10.1007/978-3-319-41528-4\_4},
  doi          = {10.1007/978-3-319-41528-4\_4},
  bibsource    = {dblp computer science bibliography, https://dblp.org}
}

@article{DBLP:journals/pacmpl/MoosbruggerSBK22,
  author       = {Marcel Moosbrugger and
                  Miroslav Stankovic and
                  Ezio Bartocci and
                  Laura Kov{\'{a}}cs},
  title        = {This is the moment for probabilistic loops},
  journal      = {Proc. {ACM} Program. Lang.},
  volume       = {6},
  number       = {{OOPSLA2}},
  pages        = {1497--1525},
  year         = {2022},
  url          = {https://doi.org/10.1145/3563341},
  doi          = {10.1145/3563341},
  bibsource    = {dblp computer science bibliography, https://dblp.org}
}

@article{DBLP:journals/pacmpl/ZaiserMO25,
  author       = {Fabian Zaiser and
                  Andrzej S. Murawski and
                  C.{-}H. Luke Ong},
  title        = {Guaranteed Bounds on Posterior Distributions of Discrete Probabilistic
                  Programs with Loops},
  journal      = {Proc. {ACM} Program. Lang.},
  volume       = {9},
  number       = {{POPL}},
  pages        = {1104--1135},
  year         = {2025},
  url          = {https://doi.org/10.1145/3704874},
  doi          = {10.1145/3704874},
  bibsource    = {dblp computer science bibliography, https://dblp.org}
}

@inproceedings{koat,
	author       = {Fabian Meyer and
	Marcel Hark and
	J{\"{u}}rgen Giesl},
	editor       = {Jan Friso Groote and
	Kim Guldstrand Larsen},
	title        = {Inferring Expected Runtimes of Probabilistic Integer Programs Using
	Expected Sizes},
	booktitle    = {Tools and Algorithms for the Construction and Analysis of Systems
	- 27th International Conference, {TACAS} 2021, Held as Part of the
	European Joint Conferences on Theory and Practice of Software, {ETAPS}
	2021, Luxembourg City, Luxembourg, March 27 - April 1, 2021, Proceedings,
	Part {I}},
	series       = {Lecture Notes in Computer Science},
	volume       = {12651},
	pages        = {250--269},
	publisher    = {Springer},
	year         = {2021},
	url          = {https://doi.org/10.1007/978-3-030-72016-2\_14},
	doi          = {10.1007/978-3-030-72016-2\_14},
	bibsource    = {dblp computer science bibliography, https://dblp.org}
}

@inproceedings{amber,
	author       = {Marcel Moosbrugger and
	Ezio Bartocci and
	Joost{-}Pieter Katoen and
	Laura Kov{\'{a}}cs},
	editor       = {Marieke Huisman and
	Corina S. Pasareanu and
	Naijun Zhan},
	title        = {The Probabilistic Termination Tool Amber},
	booktitle    = {Formal Methods - 24th International Symposium, {FM} 2021, Virtual
	Event, November 20-26, 2021, Proceedings},
	series       = {Lecture Notes in Computer Science},
	volume       = {13047},
	pages        = {667--675},
	publisher    = {Springer},
	year         = {2021},
	url          = {https://doi.org/10.1007/978-3-030-90870-6\_36},
	doi          = {10.1007/978-3-030-90870-6\_36},
	bibsource    = {dblp computer science bibliography, https://dblp.org}
}

@inproceedings{DBLP:conf/esop/HaaseBGKKKW26,
  author       = {Darion Haase and
                  Kevin Batz and
                  Adrian Gallus and
                  Benjamin Lucien Kaminski and
                  Joost{-}Pieter Katoen and
                  Lutz Klinkenberg and
                  Tobias Winkler},
  editor       = {Robbert Krebbers},
  title        = {Generating Functions Meet Occupation Measures: Invariant Synthesis
                  for Probabilistic Loops},
  booktitle    = {Programming Languages and Systems - 35th European Symposium on Programming,
                  {ESOP} 2026, Held as Part of the International Joint Conferences on
                  Theory and Practice of Software, {ETAPS} 2026, Turin, Italy, April
                  11-16, 2026, Proceedings, Part {I}},
  series       = {Lecture Notes in Computer Science},
  pages        = {314--343},
  publisher    = {Springer},
  year         = {2026},
  url          = {https://doi.org/10.1007/978-3-032-22720-1\_12},
  doi          = {10.1007/978-3-032-22720-1\_12},
  bibsource    = {dblp computer science bibliography, https://dblp.org}
}

@inproceedings{DBLP:conf/atva/HuangDM21,
  author       = {Zixin Huang and
                  Saikat Dutta and
                  Sasa Misailovic},
  editor       = {Zhe Hou and
                  Vijay Ganesh},
  title        = {{AQUA:} Automated Quantized Inference for Probabilistic Programs},
  booktitle    = {Automated Technology for Verification and Analysis - 19th International
                  Symposium, {ATVA} 2021, Gold Coast, QLD, Australia, October 18-22,
                  2021, Proceedings},
  series       = {Lecture Notes in Computer Science},
  pages        = {229--246},
  publisher    = {Springer},
  year         = {2021},
  url          = {https://doi.org/10.1007/978-3-030-88885-5\_16},
  doi          = {10.1007/978-3-030-88885-5\_16},
  bibsource    = {dblp computer science bibliography, https://dblp.org}
}

@inproceedings{batzProbabilisticProgramVerification2022,
	author       = {Kevin Batz and
	Mingshuai Chen and
	Sebastian Junges and
	Benjamin Lucien Kaminski and
	Joost{-}Pieter Katoen and
	Christoph Matheja},
	title        = {Probabilistic Program Verification via Inductive Synthesis of Inductive
	Invariants},
	booktitle    = {{TACAS} {(2)}},
	series       = {Lecture Notes in Computer Science},
	volume       = {13994},
	pages        = {410--429},
	publisher    = {Springer},
	year         = {2023},
	doi = {10.1007/978-3-031-30820-8\_25}
}

@inproceedings{DBLP:conf/cav/AndriushchenkoC21,
  author       = {Roman Andriushchenko and
                  Milan Ceska and
                  Sebastian Junges and
                  Joost{-}Pieter Katoen and
                  Simon Stupinsk{\'{y}}},
  editor       = {Alexandra Silva and
                  K. Rustan M. Leino},
  title        = {{PAYNT:} {A} Tool for Inductive Synthesis of Probabilistic Programs},
  booktitle    = {Computer Aided Verification - 33rd International Conference, {CAV}
                  2021, Virtual Event, July 20-23, 2021, Proceedings, Part {I}},
  series       = {Lecture Notes in Computer Science},
  pages        = {856--869},
  publisher    = {Springer},
  year         = {2021},
  url          = {https://doi.org/10.1007/978-3-030-81685-8\_40},
  doi          = {10.1007/978-3-030-81685-8\_40},
  bibsource    = {dblp computer science bibliography, https://dblp.org}
}

@article{DBLP:journals/pacmpl/AvanziniMS23,
  author       = {Martin Avanzini and
                  Georg Moser and
                  Michael Schaper},
  title        = {Automated Expected Value Analysis of Recursive Programs},
  journal      = {Proc. {ACM} Program. Lang.},
  volume       = {7},
  number       = {{PLDI}},
  pages        = {1050--1072},
  year         = {2023},
  url          = {https://doi.org/10.1145/3591263},
  doi          = {10.1145/3591263},
  bibsource    = {dblp computer science bibliography, https://dblp.org}
}

@article{ecoimp,
	author       = {Martin Avanzini and
	Georg Moser and
	Michael Schaper},
	title        = {A modular cost analysis for probabilistic programs},
	journal      = {Proc. {ACM} Program. Lang.},
	volume       = {4},
	number       = {{OOPSLA}},
	pages        = {172:1--172:30},
	year         = {2020},
	url          = {https://doi.org/10.1145/3428240},
	doi          = {10.1145/3428240},
	bibsource    = {dblp computer science bibliography, https://dblp.org}
}

@inproceedings{leutgebAutomatedExpectedAmortised2022,
  author       = {Lorenz Leutgeb and
                  Georg Moser and
                  Florian Zuleger},
  title        = {Automated Expected Amortised Cost Analysis of Probabilistic Data Structures},
  booktitle    = {{CAV} {(2)}},
  series       = {Lecture Notes in Computer Science},
  volume       = {13372},
  pages        = {70--91},
  publisher    = {Springer},
  year         = {2022},
  doi = {10.1007/978-3-031-13188-2_4}
}

@inproceedings{ngoBoundedExpectationsResource2018,
  title = {Bounded Expectations: Resource Analysis for Probabilistic Programs},
  shorttitle = {Bounded Expectations},
  booktitle = {Proceedings of the 39th {{ACM SIGPLAN Conference}} on {{Programming Language Design}} and {{Implementation}}},
  author = {Ngo, Van Chan and Carbonneaux, Quentin and Hoffmann, Jan},
  year = {2018},
  month = jun,
  series = {{{PLDI}} 2018},
  publisher = {{Association for Computing Machinery}},
  address = {{New York, NY, USA}},
  isbn = {978-1-4503-5698-5},
  doi = {10.1145/3192366.3192394}
}

@inproceedings{automatedterm,
	author       = {Marcel Moosbrugger and
	Ezio Bartocci and
	Joost{-}Pieter Katoen and
	Laura Kov{\'{a}}cs},
	editor       = {Nobuko Yoshida},
	title        = {Automated Termination Analysis of Polynomial Probabilistic Programs},
	booktitle    = {Programming Languages and Systems - 30th European Symposium on Programming,
	{ESOP} 2021, Held as Part of the European Joint Conferences on Theory
	and Practice of Software, {ETAPS} 2021, Luxembourg City, Luxembourg,
	March 27 - April 1, 2021, Proceedings},
	series       = {Lecture Notes in Computer Science},
	volume       = {12648},
	pages        = {491--518},
	publisher    = {Springer},
	year         = {2021},
	url          = {https://doi.org/10.1007/978-3-030-72019-3\_18},
	doi          = {10.1007/978-3-030-72019-3\_18},
	bibsource    = {dblp computer science bibliography, https://dblp.org}
}

@article{DBLP:journals/pacmpl/MajumdarS25,
  author       = {Rupak Majumdar and
                  V. R. Sathiyanarayana},
  title        = {Sound and Complete Proof Rules for Probabilistic Termination},
  journal      = {Proc. {ACM} Program. Lang.},
  volume       = {9},
  number       = {{POPL}},
  pages        = {1871--1902},
  year         = {2025},
  url          = {https://doi.org/10.1145/3704899},
  doi          = {10.1145/3704899},
  bibsource    = {dblp computer science bibliography, https://dblp.org}
}

@inproceedings{invsys1,
	author       = {Yijun Feng and
	Lijun Zhang and
	David N. Jansen and
	Naijun Zhan and
	Bican Xia},
	editor       = {Deepak D'Souza and
	K. Narayan Kumar},
	title        = {Finding Polynomial Loop Invariants for Probabilistic Programs},
	booktitle    = {Automated Technology for Verification and Analysis - 15th International
	Symposium, {ATVA} 2017, Pune, India, October 3-6, 2017, Proceedings},
	series       = {Lecture Notes in Computer Science},
	volume       = {10482},
	pages        = {400--416},
	publisher    = {Springer},
	year         = {2017},
	url          = {https://doi.org/10.1007/978-3-319-68167-2\_26},
	doi          = {10.1007/978-3-319-68167-2\_26},
	bibsource    = {dblp computer science bibliography, https://dblp.org}
}

@inproceedings{invsys2,
	author       = {Yu{-}Fang Chen and
	Chih{-}Duo Hong and
	Bow{-}Yaw Wang and
	Lijun Zhang},
	editor       = {Daniel Kroening and
	Corina S. Pasareanu},
	title        = {Counterexample-Guided Polynomial Loop Invariant Generation by Lagrange
	Interpolation},
	booktitle    = {Computer Aided Verification - 27th International Conference, {CAV}
	2015, San Francisco, CA, USA, July 18-24, 2015, Proceedings, Part
	{I}},
	series       = {Lecture Notes in Computer Science},
	volume       = {9206},
	pages        = {658--674},
	publisher    = {Springer},
	year         = {2015},
	url          = {https://doi.org/10.1007/978-3-319-21690-4\_44},
	doi          = {10.1007/978-3-319-21690-4\_44},
	bibsource    = {dblp computer science bibliography, https://dblp.org}
}

@inproceedings{invsys3,
	author       = {Joost{-}Pieter Katoen and
	Annabelle McIver and
	Larissa Meinicke and
	Carroll C. Morgan},
	editor       = {Radhia Cousot and
	Matthieu Martel},
	title        = {Linear-Invariant Generation for Probabilistic Programs: - Automated
	Support for Proof-Based Methods},
	booktitle    = {Static Analysis - 17th International Symposium, {SAS} 2010, Perpignan,
	France, September 14-16, 2010. Proceedings},
	series       = {Lecture Notes in Computer Science},
	volume       = {6337},
	pages        = {390--406},
	publisher    = {Springer},
	year         = {2010},
	url          = {https://doi.org/10.1007/978-3-642-15769-1\_24},
	doi          = {10.1007/978-3-642-15769-1\_24},
	bibsource    = {dblp computer science bibliography, https://dblp.org}
}

@inproceedings{invsys4,
	author       = {Gilles Barthe and
	Thomas Espitau and
	Luis Mar{\'{\i}}a Ferrer Fioriti and
	Justin Hsu},
	editor       = {Swarat Chaudhuri and
	Azadeh Farzan},
	title        = {Synthesizing Probabilistic Invariants via Doob's Decomposition},
	booktitle    = {Computer Aided Verification - 28th International Conference, {CAV}
	2016, Toronto, ON, Canada, July 17-23, 2016, Proceedings, Part {I}},
	series       = {Lecture Notes in Computer Science},
	volume       = {9779},
	pages        = {43--61},
	publisher    = {Springer},
	year         = {2016},
	url          = {https://doi.org/10.1007/978-3-319-41528-4\_3},
	doi          = {10.1007/978-3-319-41528-4\_3},
	bibsource    = {dblp computer science bibliography, https://dblp.org}
}

@inproceedings{cegisupast,
	author       = {Alessandro Abate and
	Mirco Giacobbe and
	Diptarko Roy},
	editor       = {Alexandra Silva and
	K. Rustan M. Leino},
	title        = {Learning Probabilistic Termination Proofs},
	booktitle    = {Computer Aided Verification - 33rd International Conference, {CAV}
	2021, Virtual Event, July 20-23, 2021, Proceedings, Part {II}},
	series       = {Lecture Notes in Computer Science},
	volume       = {12760},
	pages        = {3--26},
	publisher    = {Springer},
	year         = {2021},
	url          = {https://doi.org/10.1007/978-3-030-81688-9\_1},
	doi          = {10.1007/978-3-030-81688-9\_1},
	bibsource    = {dblp computer science bibliography, https://dblp.org}
}

@inproceedings{moments1,
	author       = {Daneshvar Amrollahi and
	Ezio Bartocci and
	George Kenison and
	Laura Kov{\'{a}}cs and
	Marcel Moosbrugger and
	Miroslav Stankovic},
	editor       = {Gagandeep Singh and
	Caterina Urban},
	title        = {Solving Invariant Generation for Unsolvable Loops},
	booktitle    = {Static Analysis - 29th International Symposium, {SAS} 2022, Auckland,
	New Zealand, December 5-7, 2022, Proceedings},
	series       = {Lecture Notes in Computer Science},
	volume       = {13790},
	pages        = {19--43},
	publisher    = {Springer},
	year         = {2022},
	url          = {https://doi.org/10.1007/978-3-031-22308-2\_3},
	doi          = {10.1007/978-3-031-22308-2\_3},
	bibsource    = {dblp computer science bibliography, https://dblp.org}
}

@inproceedings{batzPrIC3PropertyDirected2020,
  author       = {Kevin Batz and
                  Sebastian Junges and
                  Benjamin Lucien Kaminski and
                  Joost{-}Pieter Katoen and
                  Christoph Matheja and
                  Philipp Schr{\"{o}}er},
  title        = {PrIC3: Property Directed Reachability for MDPs},
  booktitle    = {{CAV} {(2)}},
  series       = {Lecture Notes in Computer Science},
  volume       = {12225},
  pages        = {512--538},
  publisher    = {Springer},
  year         = {2020},
  doi          = {10.1007/978-3-030-53291-8\_27},
}

@article{rankingsuper,
	author       = {Sheshansh Agrawal and
	Krishnendu Chatterjee and
	Petr Novotn{\'{y}}},
	title        = {Lexicographic ranking supermartingales: an efficient approach to termination
	of probabilistic programs},
	journal      = {Proc. {ACM} Program. Lang.},
	volume       = {2},
	number       = {{POPL}},
	pages        = {34:1--34:32},
	year         = {2018},
	url          = {https://doi.org/10.1145/3158122},
	doi          = {10.1145/3158122},
	bibsource    = {dblp computer science bibliography, https://dblp.org}
}

@inproceedings{kozenProbabilisticPDL1983,
  author       = {Dexter Kozen},
  title        = {A Probabilistic {PDL}},
  booktitle    = {{STOC}},
  pages        = {291--297},
  publisher    = {{ACM}},
  year         = {1983},
  doi = {10.1145/800061.808758}
}

@article{kozenProbabilisticPDL1985,
  author       = {Dexter Kozen},
  title        = {A Probabilistic {PDL}},
  journal      = {J. Comput. Syst. Sci.},
  volume       = {30},
  number       = {2},
  pages        = {162--178},
  year         = {1985},
  doi = {10.1016/0022-0000(85)90012-1}
}

@inproceedings{kaminskiWeakestPreconditionReasoning2016,
  title = {Weakest {{Precondition Reasoning}} for {{Expected Run}}\textendash{{Times}} of {{Probabilistic Programs}}},
  booktitle = {Programming {{Languages}} and {{Systems}}},
  author = {Kaminski, Benjamin Lucien and Katoen, Joost-Pieter and Matheja, Christoph and Olmedo, Federico},
  editor = {Thiemann, Peter},
  year = {2016},
  series = {Lecture {{Notes}} in {{Computer Science}}},
  publisher = {{Springer}},
  address = {{Berlin, Heidelberg}},
  isbn = {978-3-662-49498-1},
  doi = {10.1007/978-3-662-49498-1\_15}
}

@article{ProbabilisticGuardedCommands2005,
  title = {Probabilistic {{Guarded Commands Mechanized}} in {{HOL}}},
  author = {Hurd, J. and McIver, Annabelle and Morgan, Carroll},
  year = {2005},
  journal = {Theoretical Computer Science},
  volume = {346},
  number = {1},
  pages = {96--112},
  doi = {10.1016/j.tcs.2005.08.005}
}

@inproceedings{hoelzlert,
	author       = {Johannes H{\"{o}}lzl},
	editor       = {Jasmin Christian Blanchette and
	Stephan Merz},
	title        = {Formalising Semantics for Expected Running Time of Probabilistic Programs},
	booktitle    = {Interactive Theorem Proving - 7th International Conference, {ITP}
	2016, Nancy, France, August 22-25, 2016, Proceedings},
	series       = {Lecture Notes in Computer Science},
	volume       = {9807},
	pages        = {475--482},
	publisher    = {Springer},
	year         = {2016},
	url          = {https://doi.org/10.1007/978-3-319-43144-4\_30},
	doi          = {10.1007/978-3-319-43144-4\_30},
	bibsource    = {dblp computer science bibliography, https://dblp.org}
}

@book{isbabelle,
	author       = {Tobias Nipkow and
	Lawrence C. Paulson and
	Markus Wenzel},
	title        = {Isabelle/HOL - {A} Proof Assistant for Higher-Order Logic},
	series       = {Lecture Notes in Computer Science},
	volume       = {2283},
	publisher    = {Springer},
	year         = {2002},
	url          = {https://doi.org/10.1007/3-540-45949-9},
	doi          = {10.1007/3-540-45949-9},
	isbn         = {3-540-43376-7},
	bibsource    = {dblp computer science bibliography, https://dblp.org}
}

\end{document}